\documentclass[aps,pre, showpacs,amsmath,amsfonts,amssymb,reprint,groupedaddress]{revtex4-1}
\usepackage{graphicx}
\usepackage{color}
\begin{document}
\title{Entropy production in full phase space for continuous stochastic dynamics}
\author{Richard E. Spinney and
Ian J. Ford}
\affiliation{Department of Physics and Astronomy, UCL, Gower Street,
London WC1E 6BT, UK
}
\affiliation{London Centre for Nanotechnology, 17-19 Gordon Street
London WC1H 0AH, UK
}

\date{\today}
\begin{abstract}
The total entropy production and its three constituent components are described both as fluctuating trajectory-dependent quantities and as averaged contributions in the context of the continuous Markovian dynamics, described by stochastic differential equations with multiplicative noise, of systems with both odd and even coordinates with respect to time reversal, such as dynamics in full phase space. Two of these constituent quantities obey integral fluctuation theorems and are thus rigorously positive in the mean by Jensen's inequality. The third, however, is not and furthermore cannot be uniquely associated with irreversibility arising from relaxation, nor with the breakage of detailed balance brought about by non-equilibrium constraints. The properties of the various contributions to total entropy production are explored through the consideration of two examples: steady state heat conduction due to a temperature gradient, and transitions between stationary states of drift-diffusion on a ring, both in the context of the full phase space dynamics of a single Brownian particle.
\end{abstract}
\maketitle

\section{Introduction}

The concept of entropy was introduced over 150 years ago to provide a measure of the evident irreversibility of macroscopic thermodynamic phenomena. The conflict between its monotonic increase and the underlying time reversal symmetry of the microscopic dynamics, first pointed out by Loschmidt, is but one of its apparent mysteries. Nevertheless, in recent years significant insights into the nature of irreversibility and entropy production have emerged, partly due to the need for a framework to interpret thermodynamic processes on the nanoscale. These developments had their beginnings in the dissipation function and Fluctuation Theorem in deterministic thermostatted systems considered by Evans et al \cite{Evans93,Evans95,Evans02,Carberry04}, and have continued with similar concepts within the realms of chaos theory \cite{Gallavotti95} and of stochastic dynamical modelling \cite{kurchan,GCforstochastic}. Some powerful results such as the Crooks and Jarzynski relations stand out \cite{Jarzynski97,crooksoriginal,Crooks98} along with a unifying framework for overdamped Langevin dynamics \cite{seifertoriginal} based on a stochastic description of the first law of thermodynamics commonly referred to as stochastic energetics \cite{sekimoto}. In short, entropy production is a measure of the relative likelihoods of forward and reversed behaviour within the context of a model dynamical framework that includes specific dissipative terms. There are certain differences in viewpoint, but the central insight is that a mechanical (or maybe dynamical) quantity can be defined that matches the behaviour of thermodynamic entropy, in particular that on average, it increases with time. Its fluctuating nature provides additional insight into the behaviour of small systems.\\
\\
More recently, it was proposed that the entropy production associated with non-equilibrium states of small systems, arising from an underlying stochastic model of the dynamics, could be divided into two components, one related to relaxation (sometimes restricted to transitions between stationary states), and the other to any fundamental constraint that maintains the system away from an equilibrium \cite{hatanosasa,IFThousekeeping,Jarpathintegral,Esposito07,Ge09,Ge10,adiabaticnonadiabatic0,adiabaticnonadiabatic1,adiabaticnonadiabatic2}. The two components, termed adiabatic and non-adiabatic production rates respectively, were mapped onto earlier concepts known as excess and house-keeping heat transfers \cite{oono,adiabaticnonadiabatic0}. In a recent development \cite{prl}, however, it was shown that a third component of entropy production could be conceived, arising from the non-equilibrium constraint, but associated with relaxation towards the stationary state. It only arises when odd dynamical variables play a role in the dynamics, and even then only in specific cases. Only two of the three components of entropy production satisfy an integral fluctuation theorem (IFT), making them rigorously positive in the mean, properties shared by the sum of all three; the third, however, does not satisfy an IFT, and in the mean can take either sign.\\
\\
In this paper, we develop these ideas further, within a framework of full phase space continuous dynamics modelled by stochastic differential equations, with the aim of pinning down the specific form of the three contributions, both in the mean and in fluctuations about the mean and go on to make use of the formalism in some instructive example systems.

\section{Three contributions to entropy production}
We begin by considering the dynamics of a general set of variables $\textbf{\em x}=(x_1,x_2,
\ldots x_n)$ that may be odd or even under time reversal by considering the operation
$\boldsymbol{\varepsilon}\textbf{\em x}=(\varepsilon_1x_1,\varepsilon_2x_2,
\ldots \varepsilon_nx_n)$ where $\varepsilon_i=\pm 1$ for even and odd
variables $x_i$ respectively. Specifically, we consider  continuous Markovian dynamics described by a system of arbitrary uncorrelated Ito stochastic differential equations (SDEs) such that the evolution of the coordinates $\textbf{\em x}$ are given as
\begin{equation}
  dx_i=A_i(\textbf{\em x},t)dt+B_i(\textbf{\em x},t)dW_i,
\end{equation}
where $dW_i$ denotes the Wiener process. Since we allow $x_i$ to be either odd or even under time reversal we can divide the deterministic dynamics into reversible and irreversible components such that
\begin{equation}
  dx_i=A^{{\rm rev}}_i(\textbf{\em x},t)dt+A^{{\rm ir}}_i(\textbf{\em x},t)dt+B_i(\textbf{\em x},t)dW_i
\end{equation}
by defining
\begin{equation}
  A^{{\rm ir}}_i(\textbf{\em x},t)=\frac{1}{2}\left(A_i(\textbf{\em x},t)+\varepsilon_iA_i(\boldsymbol{\varepsilon}\textbf{\em x})\right)=\varepsilon_i A^{{\rm ir}}_i(\boldsymbol{\varepsilon}\textbf{\em x},t)
  \label{irprop}
\end{equation}
\begin{equation}
  A^{{\rm rev}}_i(\textbf{\em x},t)=\frac{1}{2}\left(A_i(\textbf{\em x},t)-\varepsilon_iA_i(\boldsymbol{\varepsilon}\textbf{\em x})\right)=-\varepsilon_i A^{{\rm rev}}_i(\boldsymbol{\varepsilon}\textbf{\em x},t).
  \label{revprop}
\end{equation}
 We briefly note that we intend our notation, $A(\boldsymbol{\varepsilon}\textbf{\em x},t)$, to imply a time reversal of all parameters that constitute $A$ whether they be dynamical variables included in $\textbf{\em x}$ or not. For example, a term proportional to a magnetic field appearing in $A_{x}$, where $x$ is an even spatial coordinate, would form part of $A^{\rm rev}_{x}$ since magnetic fields are odd with respect to time reversal in contrast with, for example, a force, $F$, which would appear in $A^{\rm ir}_{x}$ since force is even with respect to time reversal.\\
\\
We have specified for simplicity that  all our SDEs are driven by uncorrelated noise such that we have no cross derivatives in the corresponding Fokker-Planck equation. As such we may then represent the noise strengths as diffusion coefficients
\begin{equation}
  \frac{1}{2}B_i(\textbf{\em x},t)^2=D_i(\textbf{\em x},t)
\end{equation}
that appear in a Fokker-Planck equation describing the joint probability density function of the coordinates
\begin{align}
  &\frac{\partial p(\textbf{\em x},t)}{\partial t}=\nonumber\\
&-\sum_i\frac{\partial }{\partial x_i}\left(A_i(\textbf{\em x},t)p(\textbf{\em x},t)\right)+\sum_i\frac{\partial^2 }{\partial x^2_i}\left(D_i(\textbf{\em x},t)p(\textbf{\em x},t)\right).
\end{align}
For simplicity in the later development we assume that the diffusion coefficient is symmetric with respect to time reversal such that $D(\boldsymbol{\varepsilon}\textbf{\em x})=D(\textbf{\em x})$, which puts no restriction on the dependence on even coordinates, but requires that $D(\textbf{\em x})$ is an even function of any odd coordinates.\\
\\
 It is helpful to express the Fokker-Planck equation as a continuity equation in terms of the probability density current $J(\textbf{\em x},t)$
\begin{align}
  \frac{\partial p(\textbf{\em x},t)}{\partial t}&=-\nabla \cdot J(\textbf{\em x},t)\nonumber\\
    &=-\nabla \cdot \left(J^{\rm ir}(\textbf{\em x},t)+J^{\rm rev}(\textbf{\em x},t)\right)
\end{align}
which we separate into irreversible and reversible components. These take vector form $J=(J_1,J_2,\ldots J_n)$ as do the drift and diffusion coefficients $A=(A_1,A_2,\ldots A_n)$ and $D=(D_1,D_2,\ldots D_n)$ such that
\begin{align}
  J^{\rm ir}(\textbf{\em x},t)&=A^{{\rm ir}}(\textbf{\em x},t)p(\textbf{\em x},t)-\nabla\cdot\left(D(\textbf{\em x},t)p(\textbf{\em x},t)\right)\nonumber\\
  J^{\rm rev}(\textbf{\em x},t)&=A^{{\rm rev}}(\textbf{\em x},t)p(\textbf{\em x},t).
\end{align}
Having set out the dynamics we shall be using, we now consider the general procedure for producing quantities which obey IFTs. Given an interval of duration $\tau$, such a quantity consists of a difference between the logarithmic probability density of a given trajectory under what we shall term the forward dynamics, time dependence of the dynamics (equivalent here to an external protocol) and initial distribution of starting configurations, and that of another appropriately chosen trajectory under suitable dynamics, protocol and initial distribution. We write $p[\vec{\textbf{\em x}}]$ as the probability density of the forward
trajectory or path, $\vec{\textbf{\em x}}=\textbf{\em x}(t)$ for $0\leq t\leq \tau$,
 with a probability density function of starting configurations,
$p(\textbf{\em x}(0),0)$, which acts as an initial condition for the Fokker-Planck equation introduced earlier. A quantity that obeys an IFT is then of the form
\begin{equation}
  A[\vec{\textbf{\em x}}]=\ln\left[{p[\vec{\textbf{\em x}}]}/{p^{*}[\vec{\textbf{\em x}}^{*}]}\right]
  \label{entform}
\end{equation}
where $p^{*}[\vec{\textbf{\em x}}^{*}]$ is the probability density of a path, $\vec{\textbf{\em x}}^{*}$, under chosen dynamics (with specified nature and time dependence) and initial condition. Demonstrating the adherence of such a quantity to an IFT with respect to the forward dynamics and time dependence is straightforward by the reasoning
\begin{align}
  \langle \exp{\left[-A[\vec{\textbf{\em x}}]\right]}\rangle
&= \int d\vec{\textbf{\em x}} \;p[\vec{\textbf{\em x}}]
\exp{\left[-A[\vec{\textbf{\em x}}]\right]}\nonumber\\
&=\int d\vec{\textbf{\em x}}\;p[\vec{\textbf{\em x}}]\frac{p^{*}[\vec{\textbf{\em x}}^{*}]}
{p[\vec{\textbf{\em x}}]}\nonumber\\
  &=\int{d\vec{\textbf{\em x}}^*}\;p^{*}[\vec{\textbf{\em x}}^{*}]=1.
\label{IFT}
\end{align}
Such a result requires a Jacobian of unity for the path transformation $\vec{\textbf{\em x}}\to\vec{\textbf{\em x}}^{*}$ in order for the path integrals to be equivalent (a result assured for any involutive transformation) and the requirement $p^{*}[\vec{\textbf{\em x}}^{*}]=0$ for all $p[\vec{\textbf{\em x}}]=0$ ensuring that for normalised $p$ and $p^{*}$ all possible paths under the dynamics that produce $p^{*}$ are contained within the final integral. This may be seen as a more general version of the ergodic consistency requirement \cite{Harris07} since it is a condition required for the appropriate inclusion of paths in the final integral of Eq.~(\ref{IFT}), but we do not demand that $p$ and $p^{*}$ be non-zero for all paths. Further, any quantity $A[\vec{\textbf{\em x}}]$ based on a transformation with a Jacobian of unity takes the same form irrespective of whether probabilities or probability densities are used in the construction owing to the equivalence of measure. Such a quantity is therefore constructed unambiguously and is the direct analogue of the same quantity defined in discrete space \cite{adiabaticnonadiabatic0,prl}.
The implication of the positivity in the mean,
$\langle A[\vec{\textbf{\em x}}]\rangle\geq 0$, of the quantity $A[\vec{\textbf{\em x}}]$ is assured by Jensen's inequality. We point out that the path, $\vec{\textbf{\em x}}^{*}$, and the dynamics must be carefully chosen in order to satisfy the requirements and to produce a physically meaningful quantity.
This is particularly relevant in the presence of both odd and even variables as for many physical systems the common choice of reverse path $\textbf{\em x}^{*}(t)=\textbf{\em x}(\tau\!-\!t)$ cannot be generated under the forward dynamics, and if used to define a quantity of the form in Eq.~(\ref{entform}) would render the final integral in Eq.~(\ref{IFT}) equal to zero. However, the choice $\textbf{\em x}^{*}(t)=\boldsymbol{\varepsilon}\textbf{\em x}(\tau\!-\!t)$ typically can be generated and so leads to an IFT. However, this choice of the second path, $\textbf{\em x}^{*}(t)=\boldsymbol{\varepsilon}\textbf{\em x}(\tau\!-\!t)$, is appropriate not just because of the guarantee of an IFT, but because it means the constructed quantity $A[\vec{\textbf{\em x}}]$ serves as a measure of the irreversibility of the process and thus characterises the total entropy production.
\\\\
By following the above rules, and making the definitions $\textbf{\em x}^{\dagger}(t)=\boldsymbol{\varepsilon}\textbf{\em x}(\tau\!-\!t)$, $\textbf{\em x}^{\rm R}(t)=\textbf{\em x}(\tau\!-\!t)$ and $\textbf{\em x}^{\rm T}(t)=\boldsymbol{\varepsilon}\textbf{\em x}(t)$,  we may construct dimensionless entropy changes, which are thermodynamically meaningful when multiplied by $k_B$, of the form in Eq.~(\ref{entform}):
\begin{align}
  \Delta S_{\rm tot}&=\ln{{p[\vec{\textbf{\em x}}]}}-\ln{p^{\rm R}[\vec{\textbf{\em x}}^{\rm \dagger}]}\nonumber\\
&=\ln{\frac{p(\textbf{\em x}(0),0)}{p(\textbf{\em x}(\tau),\tau)}}+\ln{\frac{p[\textbf{\em x}(\tau)|\textbf{\em x}(0)]}{p^{\rm R}[\boldsymbol{\varepsilon}\textbf{\em x}(0)|\boldsymbol{\varepsilon}\textbf{\em x}(\tau)]}}
\label{Stotorig}
\end{align}
\begin{align}
\Delta S_1&=\ln{{p[\vec{\textbf{\em x}}]}}-\ln{p^{\rm ad,R}[\vec{\textbf{\em x}}^{\rm R}]}\nonumber\\
&=\ln{\frac{p(\textbf{\em x}(0),0)}{p(\textbf{\em x}(\tau),\tau)}}+\ln{\frac{p[\textbf{\em x}(\tau)|\textbf{\em x}(0)]}{p^{\rm ad,R}[\textbf{\em x}(0)|\textbf{\em x}(\tau)]}}
\label{S1orig}
\end{align}
\begin{align}
\Delta S_2&=\ln{{p[\vec{\textbf{\em x}}]}}-\ln{p^{\rm ad}[\vec{\textbf{\em x}}^{\rm T}]}\nonumber\\
&=\ln{\frac{p(\textbf{\em x}(0),0)}{p(\textbf{\em x}(0),0)}}+\ln{\frac{p[\textbf{\em x}(\tau)|\textbf{\em x}(0)]}{p^{\rm ad}[\boldsymbol{\varepsilon}\textbf{\em x}(\tau)|\boldsymbol{\varepsilon}\textbf{\em x}(0)]}}
\label{S2orig}
\end{align}
such that the total path probability densities are divided into initial probability density distributions and conditional probability densities. In the above the label $R$ designates a reversed protocol, equivalent here to reversed time dependence in the dynamics, and `$\rm ad$' designates that the dynamics are so-called adjoint with respect to the forward dynamics, defined as the dynamics which reach the same stationary state, but with the opposite stationary current \cite{adiabaticnonadiabatic0,Harris07,Jarpathintegral}. All are expected to obey IFTs by the nature of their form. Then, by the construction $\Delta S_{\rm tot}=\Delta S_1 +\Delta S_2+\Delta S_3$, we define
\begin{align}
\Delta S_3&=\ln{{p^{\rm ad}[\vec{\textbf{\em x}}^{\rm T}]}}+\ln{{p^{\rm ad,R}[\vec{\textbf{\em x}}^{\rm R}]}}-\ln{p[\vec{\textbf{\em x}}]}-\ln{p^{\rm R}[\vec{\textbf{\em x}}^\dagger]}\nonumber\\
&=\ln{\frac{p^{\rm ad,R}[\textbf{\em x}(0)|\textbf{\em x}(\tau)]p^{\rm ad}[\boldsymbol{\varepsilon}\textbf{\em x}(\tau)|\boldsymbol{\varepsilon}\textbf{\em x}(0)]}{p[\textbf{\em x}(\tau)|\textbf{\em x}(0)]p^{\rm R}[\boldsymbol{\varepsilon}\textbf{\em x}(0)|\boldsymbol{\varepsilon}\textbf{\em x}(\tau)]}}
\label{S3orig}
\end{align}
which cannot be expressed in the form of Eq.~(\ref{entform}) and so does not obey an IFT. By following the formalism of Seifert \cite{seifertoriginal,seifertprinciples} we identify
\begin{align}
  \Delta S_{\rm tot}&=\ln{\frac{p(\textbf{\em x}(0),0)}
{p(\textbf{\em x}(\tau),\tau)}}+\int_{t=0}^{t=\tau}d\left(\frac{\Delta Q}{k_BT_{\rm env}(\textbf{\em x}(t),t)}\right)\nonumber\\
&=\Delta S_{\rm sys}+\Delta S_{\rm med},
\label{sysmed}
\end{align}
where $\Delta S_{\rm sys}$ is known as the change in dimensionless system entropy and $\Delta S_{\rm med}$ is a generalisation of the dimensionless entropy production in the environment, or medium, for an environmental temperature, $T_{\rm env}$, which we allow to be phase space or time dependent (a property we employ in example I). The integral should be interpreted as a total medium entropy change, considered as the sum of all the incremental heat transfers to separate fixed temperature heat baths to which the particle is exposed over the course of its trajectory, divided by the appropriate temperature. Of course, this reproduces the usual $\Delta S_{\rm med}=\Delta Q/k_BT_{\rm env}$ of stochastic thermodynamics \cite{seifertoriginal} when the temperature is constant. Furthermore, to make connection with a key concept in non-equilibrium thermodynamics, we may divide the heat transfer to the environment (considering one at a constant temperature for clarity without loss of generality), into the so-called excess and house-keeping heats according to the formalism of Oono and Paniconi \cite{oono} such that for a given environmental temperature $\Delta Q=\Delta Q_{\rm ex}+\Delta Q_{\rm hk}$. To align our quantities with such a formalism we associate $\Delta S_1$ with the excess heat
\begin{equation}
  \Delta Q_{\rm ex}=(\Delta S_1\!-\!\Delta S_{\rm sys})k_BT_{\rm env},
\end{equation}
$\Delta S_2$ with a so called `generalised house-keeping heat' \cite{prl}
\begin{equation}
  \Delta Q_{\rm hk,G}=\Delta S_2k_BT_{\rm env}
\end{equation}
and $\Delta S_3$ with  the `transient house-keeping heat',
\begin{equation}
  \Delta Q_{\rm hk,T}=\Delta S_3k_BT_{\rm env},
\end{equation}
named to reflect its mean behaviour, such that $\Delta Q_{\rm hk}=\Delta Q_{\rm hk,G}+\Delta Q_{\rm hk,T}$. Like previous formalisms \cite{adiabaticnonadiabatic0,adiabaticnonadiabatic2} where the entropy production was divided into two contributions associated with relaxation, and an absence of detailed balance, respectively, both in the mean and in detail \cite{adiabaticnonadiabatic0,adiabaticnonadiabatic1,adiabaticnonadiabatic2}, we have a contribution $\Delta S_1$ which is non-zero only in the presence of relaxation and $\Delta S_2$ which is non-zero only in the absence of detailed balance  both in the mean and in detail. However, we also have a quantity $\Delta S_3$ which is non-zero in detail only in the absence of detailed balance, but only contributes in the mean during the course of relaxation. Such a formalism asserts that the two origins of entropy production may often be more closely related, with such a circumstance arising under the inclusion of odd variables and when the stationary distribution is asymmetric in any of those odd variables. The aim of this paper is to derive the equations of motion for each of these quantities for continuous stochastic systems and to illustrate their behaviour through some simple examples.
\section{Representing entropy production for continuous behaviour}
Since we are describing the dynamics using SDEs it is sensible to seek a description of a small increment in each entropy production given an increment in the underlying variables $\textbf{\em x}'-\textbf{\em x}=\textbf{\em x}(t+dt)-\textbf{\em x}(t)$ in a time $dt$ so that we identify from Eqs.~(\ref{Stotorig}), (\ref{S1orig}), (\ref{S2orig}) and (\ref{S3orig})
\begin{align}
  d\Delta S_{\rm tot}&=-d(\ln{p})+\ln{\frac{p(\textbf{\em x}',t+dt|\textbf{\em x},t)}{p(\boldsymbol{\varepsilon}\textbf{\em x},t+dt|\boldsymbol{\varepsilon}\textbf{\em x}',t)}}\\
d\Delta S_{\rm 1}&=-d(\ln{p})+\ln{\frac{p(\textbf{\em x}',t+dt|\textbf{\em x},t)}{p^{\rm ad}(\textbf{\em x},t+dt|\textbf{\em x}',t)}}\\
d\Delta S_{\rm 2}&=\ln{\frac{p(\textbf{\em x}',t+dt|\textbf{\em x},t)}{p^{\rm ad}(\boldsymbol{\varepsilon}\textbf{\em x}',t+dt|\boldsymbol{\varepsilon}\textbf{\em x},t)}}\\
d\Delta S_{\rm 3}&=\ln{\frac{p^{\rm ad}(\textbf{\em x},t+dt|\textbf{\em x}',t)p^{\rm ad}(\boldsymbol{\varepsilon}\textbf{\em x}',t+dt|\boldsymbol{\varepsilon}\textbf{\em x},t)}{p(\textbf{\em x}',t+dt|\textbf{\em x},t)p(\boldsymbol{\varepsilon}\textbf{\em x},t+dt|\boldsymbol{\varepsilon}\textbf{\em x}',t)}}
\end{align}
thereby establishing the SDEs that describe entropy production and noting the abbreviation $d(\ln(p))=\ln{(p(\textbf{\em x}(t+dt),t+dt)/p(\textbf{\em x}(t),t))}$.\\
\\
To proceed we require a representation of the path probabilities in these expressions valid over the small time interval $dt$. This may be achieved by considered the short time Green's function or `short time propagator' \cite{shorttimepropagator} which is given generally as the conditional probability of a displacement $dx=x'-x$ in a time $dt$ subject to a delta function initial condition and is of the form
\begin{align}
  &p(\textbf{\em x}',t\!+\!dt|\textbf{\em x},t)=\prod_i\sqrt{\frac{1}{4\pi D_{i}(\textbf{\em r},t)dt}}\nonumber\\
  &\!\times\!\exp\!\left[\!-\!\frac{\left(dx_i\!-\!A_i(\textbf{\em r},t)dt\!+\!2a(\partial D_{i}(\textbf{\em r},t)/\partial r_i)dt\right)^2}{4D_{i}(\textbf{\em r})dt}\right.\nonumber\\
      &\left.\qquad\!-\!adt\frac{\partial A_{i}(\textbf{\em r},t)}{\partial r_i}\!+\!a^2dt\frac{\partial^2 D_{i}(\textbf{\em r},t)}{\partial r_i^2}\right]
  \label{propagator}
\end{align}
where $dx_i=x'_i\!-\!x_i$ and where $a$ is a free parameter ranging from $0$ to $1$ which defines the evaluation point of certain terms in the propagator $\textbf{\em r}=a\textbf{\em x}'\!+\!(1\!-\!a)\textbf{\em x}$ and ${r_i}=a{x_i}'\!+\!(1\!-\!a){x_i}$, and which reflects the ambiguity of a discretised interpretation of continuous stochastic behaviour. We note, however that as $dt\to 0$ all forms for the propagator are correct: they are all accurate to first order in $dt$ and result in the same Fokker-Planck equation.\\
\\
We wish to construct the increment in entropy production in the medium by a consideration of
\begin{equation}
 d\Delta S_{\rm med}=\ln{\frac{p(\textbf{\em x}',t\!+\!dt|\textbf{\em x},t)}{p(\boldsymbol{\varepsilon}\textbf{\em x},t\!+\!dt|\boldsymbol{\varepsilon}\textbf{\em x}',t)}}
  \label{lim}
\end{equation}
by employing the appropriate reverse short time propagator
\begin{widetext}
\begin{center}
\begin{align}
 &p(\boldsymbol{\varepsilon}\textbf{\em x},t+dt|\boldsymbol{\varepsilon}\textbf{\em x}',t)=\nonumber\\
&\prod_{i}\sqrt{\frac{1}{4\pi D_{i}(\boldsymbol{\varepsilon}\textbf{\em r}',t)dt}}
  \exp\left[-\frac{\left(-\varepsilon_idx_i-A_i(\boldsymbol{\varepsilon}\textbf{\em r}',t)dt+2b(\partial D_{i}(\boldsymbol{\varepsilon}\textbf{\em r}',t)/\partial (\varepsilon_ir'_i))dt\right)^2}{4D_{i}(\boldsymbol{\varepsilon}\textbf{\em r}')dt}-bdt\frac{\partial A_{i}(\boldsymbol{\varepsilon}\textbf{\em r}',t)}{\partial (\varepsilon_ir'_i)}+b^2dt\frac{\partial^2 D_{i}(\boldsymbol{\varepsilon}\textbf{\em r}',t)}{\partial (\varepsilon_ir'_i)^2}\right]
\end{align}
\end{center}
where $b$ is a corresponding free parameter ranging from $0$ to $1$  such that $\textbf{\em r}'=b\textbf{\em x}\!+\!(1\!-\!b)\textbf{\em x}'$ and ${r'_i}=b{x_i}\!+\!(1\!-\!b){x'_i}$. Using  Eqs.~(\ref{irprop}) and (\ref{revprop}) along with the assumption $D_i(\boldsymbol{\varepsilon}\textbf{\em x})=D_i(\textbf{\em x})$ such that any diffusion constants for odd variables are symmetric, we may write
\begin{center}
\begin{align}
  p(\boldsymbol{\varepsilon}\textbf{\em x},t+dt|\boldsymbol{\varepsilon}\textbf{\em x}',t)&=\prod_{i}\sqrt{\frac{1}{4\pi D_{i}(\textbf{\em r}',t)dt}}\exp\left[-\frac{\left(-\varepsilon_idx_i-\varepsilon_i\left(-A_i^{\rm rev}(\textbf{\em r}',t)+A^{\rm ir}_i(\textbf{\em r}',t)\right)dt+2b(\partial D_{i}(\textbf{\em r}',t)/\partial (\varepsilon_ir'_i))dt\right)^2}{4D_{i}(\textbf{\em r}')dt}\right.\nonumber\\
&\qquad\qquad\left.-bdt\left(\frac{\partial \varepsilon_iA^{\rm ir}_{i}(\textbf{\em r}',t)}{\partial (\varepsilon_ir'_i)}-\frac{\partial \varepsilon_iA^{\rm rev}_{i}(\textbf{\em r}',t)}{\partial (\varepsilon_ir'_i)}\right)+b^2dt\frac{\partial^2  D_{i}(\textbf{\em r}',t)}{\partial (\varepsilon_i r'_i)^2}\right]
\end{align}
\end{center}
which is the same as
\begin{center}
\begin{align}
  p(\boldsymbol{\varepsilon}\textbf{\em x},t+dt|\boldsymbol{\varepsilon}\textbf{\em x}',t)&=\prod_{i}\sqrt{\frac{1}{4\pi D_{i}(\textbf{\em r}',t)dt}}
  \exp\left[-\frac{\left(-dx_i-\left(-A_i^{\rm rev}(\textbf{\em r}',t)+A^{\rm ir}_i(\textbf{\em r}',t)\right)dt+2b(\partial D_{i}(\textbf{\em r}',t)/\partial r'_i)dt\right)^2}{4D_{i}(\textbf{\em r}')dt}\right.\nonumber\\
&\qquad\qquad\left.-bdt\left(\frac{\partial A^{\rm ir}_{i}(\textbf{\em r}',t)}{\partial r'_i}-\frac{\partial A^{\rm rev}_{i}(\textbf{\em r}',t)}{\partial r'_i}\right)+b^2dt\frac{\partial^2 D_{i}(\textbf{\em r}',t)}{\partial {r'}_i^2}\right].
\label{backprop}
\end{align}
\end{center}
\end{widetext}
The mathematical details necessary for the development of Eq.~(\ref{lim}), which due to their somewhat cumbersome nature we leave to appendix \ref{appendixA}, reveal that for multiplicative noise one obtains a result which is dependent on the choice $a$ and $b$. The resolution of this apparent arbitrariness is not related to the nature of the underlying SDEs, but rather on consistently using the equivalent evaluation point for forward and time reversed paths on an infinitesimal scale. The normal rules of calculus would dictate no dependence, but different rules apply to SDEs and stochastic calculus. Such a consideration reveals the correct choice in Eq.~(\ref{backprop}) to be $b=1-a$ with $a$ remaining as a free parameter. This yields the unambiguous Ito SDE for the medium entropy change
\begin{align}
  d\Delta S_{\rm med}&=\sum_{i}\frac{A_i^{\rm ir}(\textbf{\em x})}{D_{i}(\textbf{\em x})}dx_i-\frac{A^{\rm rev}_i(\textbf{\em x})A^{\rm ir}_i(\textbf{\em x})}{D_{i}(\textbf{\em x})}dt\nonumber\\
&+\frac{\partial A_i^{\rm ir}(\textbf{\em x})}{\partial x_i}dt-\frac{\partial A_i^{\rm rev}(\textbf{\em x})}{\partial x_i}dt\nonumber\\
&-\frac{1}{D_{i}(\textbf{\em x})}\frac{\partial D_{i}(\textbf{\em x})}{\partial x_i}dx_i\nonumber\\
&+\frac{(A^{\rm rev}(\textbf{\em x})-A^{\rm ir}(\textbf{\em x}))}{D_{i}(\textbf{\em x})}\frac{\partial D_{i}(\textbf{\em x})}{\partial x_i}dt\nonumber\\
&-\frac{\partial^2 D_{i}(\textbf{\em x})}{\partial x_i^2}dt+\frac{1}{D_{i}(\textbf{\em x})}\left(\frac{\partial D_{i}(\textbf{\em x})}{\partial x_i}\right)^2dt
\label{itomed}
\end{align}
where for brevity we use notation $f(\textbf{\em x})\equiv f(\textbf{\em x},t)$. To clarify, in such an approach choices may include Stratonovich ($a=b=1/2$) evaluation for both propagators in Eq.~(\ref{lim}), a choice which is implicitly used by many authors \cite{Jarpathintegral,seifertprinciples} within integrated Onsager-Machlup approaches, but does not preclude others in the construction of SDEs such as, for example, an Ito prescription ($a=0$) in the forward propagator and a Hanggi-Klimontovich ($b=1$) in the backwards propagator. We point out that all evaluation points lead to the correct \emph{path probability} when supplemented with the correct multiplication scheme, but that if one has multiplicative noise, the correct representation of the \emph{entropy production} requires the more exact relation between the evaluation points.\\
\\
 Proceeding, we may now construct an SDE for the total entropy production by first considering an increment in the system entropy which under Ito rules is
\begin{align}
 d\Delta S_{\rm sys}&=- d(\ln{p(\textbf{\em x})})\nonumber\\
&=-\frac{1}{p(\textbf{\em x})}\frac{\partial p(\textbf{\em x})}{\partial t}dt-\frac{1}{p(\textbf{\em x})}\sum_i\frac{\partial p(\textbf{\em x})}{\partial x_i}dx_i\nonumber\\
&-\sum_{i}\frac{D_{i}(\textbf{\em x})}{p(\textbf{\em x})}\left(\frac{\partial^2 p(\textbf{\em x})}{\partial x_i^2}-\frac{1}{p(\textbf{\em x})}\left(\frac{\partial p(\textbf{\em x})}{\partial x_i}\right)^2\right)dt,
\end{align}
which together with Eq.~(\ref{itomed}), and after insertion of the Fokker-Planck equation, leads to
\begin{align}
 d\Delta S_{\rm tot}\!&=\sum_i-\frac{1}{p(\textbf{\em x})}\frac{\partial p(\textbf{\em x})}{\partial x_i}dx_i+\frac{1}{p(\textbf{\em x})} \frac{\partial (A_i(\textbf{\em x})p(\textbf{\em x}))}{\partial x_i}dt\nonumber\\
&\!\!\!\!\!\!\!\!\!\!\!\!\!-\frac{1}{p(\textbf{\em x})}\left(\frac{\partial^2(D_i(\textbf{\em x})p(\textbf{\em x}))}{\partial x^2_i}+D_i(\textbf{\em x})\frac{\partial^2 p(\textbf{\em x})}{\partial x_i^2}\right.\nonumber\\
  &\left.\qquad\qquad-\frac{D_i(\textbf{\em x})}{p(\textbf{\em x})}\left(\frac{\partial p(\textbf{\em x})}{\partial x_i}\right)^2\right)dt\nonumber\\
&+\frac{A_i^{\rm ir}(\textbf{\em x})}{D_{i}(\textbf{\em x})}dx_i-\frac{A^{\rm rev}_i(\textbf{\em x})A^{\rm ir}_i(\textbf{\em x})}{D_{i}(\textbf{\em x})}dt\nonumber\\
&+\frac{\partial A_i^{\rm ir}(\textbf{\em x})}{\partial x_i}dt-\frac{\partial A_i^{\rm rev}(\textbf{\em x})}{\partial x_i}dt\nonumber\\
&-\frac{1}{D_{i}(\textbf{\em x})}\frac{\partial D_{i}(\textbf{\em x})}{\partial x_i}dx_i\nonumber\\
&+\frac{(A^{\rm rev}(\textbf{\em x})-A^{\rm ir}(\textbf{\em x}))}{D_{i}(\textbf{\em x})}\frac{\partial D_{i}(\textbf{\em x})}{\partial x_i}dt\nonumber\\
&-\frac{\partial^2 D_{i}(\textbf{\em x})}{\partial x_i^2}dt+\frac{1}{D_{i}(\textbf{\em x})}\left(\frac{\partial D_{i}(\textbf{\em x})}{\partial x_i}\right)^2dt.
\end{align}
If Stratonovich rules, for example, are preferred we can write (by definition of the Stratonovich integral, indicated by the $\circ$ notation)
\begin{align}
  d\Delta S_{\rm tot}&=\sum_i-\frac{1}{p(\textbf{\em x})}\frac{\partial p(\textbf{\em x})}{\partial x_i}\circ dx_i+\frac{1}{p(\textbf{\em x})} \frac{\partial (A_i(\textbf{\em x})p(\textbf{\em x}))}{\partial x_i}dt\nonumber\\
&-\frac{1}{p(\textbf{\em x})}\left(\frac{\partial^2(D_i(\textbf{\em x})p(\textbf{\em x}))}{\partial x^2_i}\right)dt\nonumber\\
&+\frac{A_i^{\rm ir}(\textbf{\em x})}{D_{i}(\textbf{\em x})}\circ dx_i-\frac{A^{\rm rev}_i(\textbf{\em x})A^{\rm ir}_i(\textbf{\em x})}{D_{i}(\textbf{\em x})}dt\nonumber\\
&-D_{i}(\textbf{\em x})\frac{\partial}{\partial x_i}\left(\frac{A^{\rm rev}_i(\textbf{\em x})}{D_{i}(\textbf{\em x})}\right)dt\nonumber\\
&-\frac{1}{D_{i}(\textbf{\em x})}\frac{\partial D_{i}(\textbf{\em x})}{\partial x_i}\circ dx_i.
\label{StotSDE}
\end{align}
This is a very general and robust definition of the entropy production for continuous stochastic behaviour and can be thought of as a generalisation of the pioneering approach in \cite{seifertoriginal}  wherein the equation of motion for entropy essentially describes $d\Delta S_{\rm tot}$ for a specific system with additive noise, even variables ($\boldsymbol{\varepsilon}\textbf{\em x}=\textbf{\em x}$) and implicitly using Stratonovich rules.\\
\\
 We point out that such a construction allows us to consider purely deterministic coordinates ($D_i(\textbf{\em x})=0$) as would apply, for example, to the case of spatial coordinates within a full phase space Langevin description. In such coordinates $D_i(\textbf{\em x})$ is assumed constant and taken to zero. The remaining terms then clearly diverge unless we demand $A_i^{\rm ir}(\textbf{\em x})=0$ since in these instances, for the reverse path to be a solution to the forward dynamics the motion must be purely reversible. This condition simply amounts to the requirement that the reverse path exists. There is, however, a contribution to the medium entropy production, due to the dynamics of these coordinates, technically since path probability densities, not probabilities, are being considered in the formulation. The contribution to the medium entropy production due to the deterministic behaviour of these coordinates is
\begin{equation}
  \Delta S_{\rm med,det}=-\frac{\partial A_i^{\rm rev}(\textbf{\em x})}{\partial x_i}dt,
\end{equation}
a result that provides an insight into the similarities and differences between stochastic and deterministic measures of irreversibility: it is demonstrably equal to the phase space contraction found in non-linear dynamical systems, which is associated with the heat transfer to the environment brought about by thermostatting terms in such approaches. This leads to a quantity that is positive in the mean for deterministic systems: the dissipation function \cite{Evans95}. We point out, however that \emph{total entropy production}, as defined here for stochastic systems, is zero for deterministic dynamics. This is because the change in the system entropy would be equal and opposite to the change in medium entropy, technically since it involves probability densities at the start and end of the process. In contrast the dissipation function can provide a measure of irreversibility because it involves a comparison of trajectories originating \emph{from the same} starting distribution. This contrast is to be expected as the total entropy production, as defined for the systems we consider, arises from explicit irreversibility in the dynamics, which deterministic, reversible equations do not provide.
\\\\
Frequently the average entropy production rate is argued to be proportional to the mean probability flux squared, as derived, for example, by taking the time derivative of the Gibbs entropy of a system, and identifying an evidently positive contribution as the total entropy production rate and the remainder as the (negative of) the medium entropy production rate \cite{adiabaticnonadiabatic2,Tome}. We prefer however, to derive the average contributions directly from the SDEs so that we can avoid arbitrarily identifying a positive contribution with a quantity expected to obey an IFT: strictly there is no guarantee such a division is unique, as another description shows \cite{positivewrong}. To do so is straightforward and requires us to find the average increment in $\Delta S_{\rm tot}$ by means of the integral
\begin{equation}
 \langle d \Delta S_{\rm tot}\rangle= \int d\textbf{\em x}\int d\textbf{\em x}'\;p(\textbf{\em x},t)p(\textbf{\em x}',t+dt|\textbf{\em x},t) d\Delta S_{\rm tot}.
\label{average}
\end{equation}
The benefit of such a formulation is that we may characterise $d \Delta S_{\rm tot}$ using an Ito SDE based on the underlying relations $dx_i=A_idt+B_idW_i$ and then use the martingale property of the Ito stochastic integral $\langle B_idW_i \rangle=0$ since $B_i$ is non-anticipating, such that
we can simplify the integral in Eq.~(\ref{average}) by writing
\begin{equation}
 \langle d \Delta S_{\rm tot}\rangle= \int d\textbf{\em x}\;p(\textbf{\em x})\langle d\Delta S_{\rm tot}|\textbf{\em x}\rangle
\end{equation}
and evaluating the conditional average $\langle d\Delta S_{\rm tot}|\textbf{\em x}\rangle$ by replacing all occurrences of $dx_i$ with $(A_i^{\rm ir}+A_i^{\rm rev})dt$ in $d\Delta S_{\rm tot}$. We thus get
\begin{align}
  \langle d\Delta S_{\rm tot}\rangle&=\sum_{i}\left[\int d{\textbf{\em x}} \;\frac{p(\textbf{\em x})(A^{\rm ir}(\textbf{\em x}))^2}{D_{i}(\textbf{\em x})}+2p(\textbf{\em x})\frac{\partial A^{\rm ir}_i(\textbf{\em x})}{\partial x_i}\right.\nonumber\\
&\left.-2p(\textbf{\em x})\frac{A^{\rm ir}(\textbf{\em x})}{D_{i}(\textbf{\em x})}\frac{\partial D_{i}(\textbf{\em x})}{\partial x_i}-p(\textbf{\em x})\frac{\partial^2 D_{i}(\textbf{\em x})}{\partial x_i^2}\right.\nonumber\\
&\left.+\frac{p(\textbf{\em x})}{D_{i}(\textbf{\em x})}\left(\frac{\partial D_{i}(\textbf{\em x})}{\partial x_i}\right)^2 -\frac{\partial^2 D_{i}(\textbf{\em x})p(\textbf{\em x})}{\partial x_i^2}\right.\nonumber\\
&\left.-D_{i}(\textbf{\em x})\frac{\partial^2 p(\textbf{\em x})}{\partial x_i^2}+\frac{D_{i}(\textbf{\em x})}{p(\textbf{\em x})}\left(\frac{\partial p(\textbf{\em x})}{\partial x_i}\right)^2\right]dt.
\end{align}
By applying the product rule, integrating by parts and assuming $p(\textbf{\em x})$ and $\partial p(\textbf{\em x})/\partial x_i$ either vanish or cancel at the boundaries, we may simplify to find the total entropy production rate
 \begin{align}
&\frac{d\langle \Delta S_{\rm tot}\rangle}{dt}=\nonumber\\
&\sum_{i}\int d\textbf{\em x}\; \frac{\left(p(\textbf{\em x})A^{\rm ir}_i(\textbf{\em x})-D_{i}(\textbf{\em x})\frac{\partial p(\textbf{\em x})}{\partial x_i}-p(\textbf{\em x})\frac{\partial D_{i}(\textbf{\em x})}{\partial x_i}\right)^2}{p(\textbf{\em x})D_{i}(\textbf{\em x})}
\end{align}
or more concisely
\begin{equation}
\frac{d\langle \Delta S_{\rm tot}\rangle}{dt}=\sum_{i}\int d\textbf{\em x}\;\frac{(J^{\rm ir}_i(\textbf{\em x}))^2}{p(\textbf{\em x})D_i(\textbf{\em x})},
\label{StotAV}
\end{equation}
providing an expression for the mean instantaneous entropy production rate which is rigorously positive, as it must be because of the adherence of $\Delta S_{\rm tot}$ to an IFT, and is dependent on the \emph{irreversible} flux.

\section{Expressions for $\Delta S_1$, $\Delta S_2$ and $\Delta S_3$}
In order to consider a division of the entropy production into the thermodynamically meaningful quantities outlined above we are required to construct path probabilities using the so-called adjoint dynamics. These dynamics may not be physically realisable: for example they may require negative positional steps to result from positive velocities (as indicated by the paths $\textbf{\em x}^{\rm R}(t)$ and $\textbf{\em x}^{\rm T}(t)$), but this is of no concern since they are only introduced for the mathematical construction of the entropy contributions. We consider an arbitrary stationary distribution of a given system which may be written in terms of a non-equilibrium potential, $\phi(\textbf{\em x})$, such that
\begin{equation}
  p^{\rm st}(\textbf{\em x})=\exp[-\phi(\textbf{\em x})]
\label{NEP}
\end{equation}
and assert that the adjoint dynamics are those that result in the same stationary distribution, but have an opposite flux. As such we require
\begin{equation}
  \frac{\partial p^{\rm st}(\textbf{\em x})}{\partial t}=-\nabla \cdot J^{\rm st}(\textbf{\em x})=\nabla \cdot J^{\rm st,ad}(\textbf{\em x})=0
\end{equation}
with
\begin{equation}
  J^{\rm st,ad}(\textbf{\em x})=-J^{\rm st}(\textbf{\em x}).
\end{equation}
In order to characterise the adjoint dynamics we construct the adjoint flux according to
\begin{align}
  J^{\rm st,ad}(\textbf{\em x})&=A^{{\rm ad}}(\textbf{\em x})p^{\rm st}(\textbf{\em x})-\frac{\partial }{\partial x_i}\left(D_i(\textbf{\em x})p^{\rm st}(\textbf{\em x})\right)\nonumber\\
&=A^{{\rm ad}}(\textbf{\em x})e^{-\phi(\textbf{\em x})}-\frac{\partial }{\partial x_i}\left(D_i(\textbf{\em x})e^{-\phi(\textbf{\em x})}\right)\nonumber\\
&=\left(A^{{\rm ad}}(\textbf{\em x})-\frac{\partial D_i(\textbf{\em x})}{\partial x_i}+D_i(\textbf{\em x})\frac{\partial \phi(\textbf{\em x})}{\partial x_i}\right)e^{-\phi(\textbf{\em x})}\nonumber\\
&=-\left(A(\textbf{\em x})-\frac{\partial D_i(\textbf{\em x})}{\partial x_i}+D_i(\textbf{\em x})\frac{\partial \phi(\textbf{\em x})}{\partial x_i}\right)e^{-\phi(\textbf{\em x})}.
\end{align}
Consequently we have the requirement
\begin{equation}
   A^{{\rm ad}}(\textbf{\em x})=-A(\textbf{\em x})+2\frac{\partial D_i(\textbf{\em x})}{\partial x_i}-2D_i(\textbf{\em x})\frac{\partial \phi(\textbf{\em x})}{\partial x_i}.
   \label{adjointA}
\end{equation}
Let us now consider the quantity
\begin{equation}
  d\Delta S_{\rm ex}=\ln{\frac{p(\textbf{\em x}',t+dt|\textbf{\em x},t)}{p^{\rm ad}(\textbf{\em x},t+dt|\textbf{\em x}',t)}},
  \label{S1part}
\end{equation}
where $\Delta S_{\rm ex}=\Delta Q_{\rm ex}/k_BT_{\rm env}$, which we have previously asserted constitutes part of the incremental contribution to the quantity $\Delta S_1$ based on relations in Eqs.~(\ref{S1orig}) and its short time representation. We evaluate Eq.~(\ref{S1part}), taking the transition probability density in the numerator from Eq.~(\ref{propagator}) and, for convenience, choosing $a=1/2$. We can represent the transition probability density appearing in the denominator through a similar construction, but using a substitution for the adjoint drift term from Eq.~(\ref{adjointA}), together with the complementary evaluation point choice $b=1-a=1/2$ such that
\begin{widetext}
\begin{center}
\begin{align}
  &p^{\rm ad}(\textbf{\em x},t+dt|\textbf{\em x}',t)=\prod_i\sqrt{\frac{1}{4\pi D_{i}(\textbf{\em r})dt}}\exp\left[-\frac{\left(-dx_i+(A_i(\textbf{\em r})-2(\partial D_i(\textbf{\em r})/\partial r_i)+2D_i(\textbf{\em r})(\partial \phi(\textbf{\em r}) /\partial r_i))dt+(\partial D_{i}(\textbf{\em r})/\partial r_i)dt\right)^2}{4D_{i}(\textbf{\em r})dt}\right.\nonumber\\
      &\left.\qquad+\frac{dt}{2}\frac{\partial }{\partial r_i}\left(A_{i}(\textbf{\em r})-2\frac{\partial D_i(\textbf{\em r})}{\partial r_i}+2D_i(\textbf{\em r})\frac{\partial \phi(\textbf{\em r}) }{\partial r_i}\right)+\frac{dt}{4}\frac{\partial^2 D_{i}(\textbf{\em r})}{\partial r_i^2}\right].
\end{align}
\end{center}
\end{widetext}
Since we have in both cases chosen evaluation at $a=b=1/2$ we note that multiplication follows Stratonovich rules so that we have $f(\textbf{\em r})dx_i=f(\textbf{\em x})\circ dx_i$. Considering the ratio of these two propagators we find
\begin{align}
&d\Delta S_{\rm ex}=\ln{\frac{p(\textbf{\em x}',t+dt|\textbf{\em x},t)}{p^{\rm ad}(\textbf{\em x},t+dt|\textbf{\em x}',t)}}=\nonumber\\
 &\sum_i D_i(\textbf{\em x})\left(\frac{\partial \phi(\textbf{\em x})}{\partial x_i}\right)^2dt+A_i(\textbf{\em x})\frac{\partial \phi(\textbf{\em x})}{\partial x_i}dt\nonumber\\
&-\frac{\partial \phi(\textbf{\em x})}{\partial x_i}\circ dx_i-\frac{\partial A_i(\textbf{\em x})}{\partial x_i}dt+\frac{\partial^2 D_i(\textbf{\em x})}{\partial x_i^2}dt\nonumber\\
&-D_i(\textbf{\em x})\frac{\partial^2\phi(\textbf{\em x})}{\partial x_i^2}dt-2\frac{\partial D_i(\textbf{\em x})}{\partial x_i}\frac{\partial \phi(\textbf{\em x})}{\partial x_i}dt.
\end{align}
However, we also have the condition
\begin{align}
  &\nabla\cdot J^{\rm st}(\textbf{\em x})=0\nonumber\\
&=\sum_i\frac{\partial }{\partial x_i}\left(e^{-\phi(\textbf{\em x})}\left(A_i(\textbf{\em x})-\frac{\partial D_i(\textbf{\em x})}{\partial x_i}+D_i(\textbf{\em x})\frac{\partial \phi(\textbf{\em x})}{\partial x_i}\right)\right)\nonumber\\
&=\left(-A_i(\textbf{\em x})\frac{\partial \phi(\textbf{\em x})}{\partial x_i}-D_i(\textbf{\em x})\left(\frac{\partial \phi(\textbf{\em x})}{\partial x_i}\right)^2+\frac{\partial A_i(\textbf{\em x})}{\partial x_i}\right.\nonumber\\
&\left.-\frac{\partial^2D_i(\textbf{\em x})}{\partial x_i^2}+D_i(\textbf{\em x})\frac{\partial^2\phi(\textbf{\em x})}{\partial x_i^2}+2\frac{\partial D_i(\textbf{\em x})}{\partial x_i}\frac{\partial\phi(\textbf{\em x})}{\partial x_i}\right)e^{-\phi(\textbf{\em x})}
\label{stationary}
\end{align}
and so by insertion we arrive at
\begin{equation}
\ln{\frac{p(\textbf{\em x}',t+dt|\textbf{\em x},t)}{p^{\rm ad}(\textbf{\em x},t+dt|\textbf{\em x}',t)}}=\sum_i-\frac{\partial \phi(\textbf{\em x})}{\partial x_i}\circ dx_i
\end{equation}
which justifies the usual characterisation of the adjoint dynamics \cite{adiabaticnonadiabatic0,Harris07,Jarpathintegral} for use in continuous dynamics when written
\begin{equation}
\frac{p(\textbf{\em x}',t+dt|\textbf{\em x},t)}{p^{\rm ad}(\textbf{\em x},t+dt|\textbf{\em x}',t)}=\frac{p^{\rm st}(\textbf{\em x}')}{p^{\rm st}(\textbf{\em x})}
\end{equation}
through consideration of Eq.~(\ref{NEP}) and the Stratonovich rules which mimic normal calculus.

We construct an increment in $\Delta S_1$, using the above result with the inclusion of a change in system entropy such that
\begin{align}
  d\Delta S_1&=d\Delta S_{\rm sys}+d\left(\Delta Q_{\rm ex}/k_BT_{\rm env}\right)\nonumber\\
&=-d(\ln p)-\sum_i\frac{\partial \phi(\textbf{\em x})}{\partial x_i}\circ dx_i\nonumber\\
  &=-\frac{1}{p(\textbf{\em x})}\frac{\partial p(\textbf{\em x})}{\partial t}dt-\sum_{i}\frac{1}{p(\textbf{\em x})}\frac{\partial p(\textbf{\em x})}{\partial x_i}dx_i\nonumber\\
&-D_{i}(\textbf{\em x})\left(\frac{1}{p(\textbf{\em x})}\frac{\partial^2 p(\textbf{\em x})}{\partial x_i^2}-\frac{1}{(p(\textbf{\em x}))^2}\left(\frac{\partial p(\textbf{\em x})}{\partial x_i}\right)^2\right)dt\nonumber\\
  &-\frac{\partial \phi(\textbf{\em x})}{\partial x_i} dx_i-D_i(\textbf{\em x})\frac{\partial^2\phi(\textbf{\em x})}{\partial x_i^2}dt.
  \label{S1SDE}
\end{align}
Applying the same averaging procedure used to calculate $\langle d\Delta S_{\rm tot}\rangle$ we find
\begin{align}
  \langle d\Delta S_1\rangle&=\sum_i\int d\textbf{\em x}\;p(\textbf{\em x})\frac{\partial A_i(\textbf{\em x})}{\partial x_i}dt-\frac{\partial^2(D_i(\textbf{\em x})p(\textbf{\em x}))}{\partial x_i^2}dt\nonumber\\
&+\frac{D_i(\textbf{\em x})}{p(\textbf{\em x})}\left(\frac{\partial p(\textbf{\em x})}{\partial x_i}\right)^2dt-D_i(\textbf{\em x})\frac{\partial^2p(\textbf{\em x})}{\partial x_i^2}dt\nonumber\\
&-p(\textbf{\em x})A(\textbf{\em x})\frac{\partial \phi(\textbf{\em x})}{\partial x_i}dt-p(\textbf{\em x})D(\textbf{\em x})\frac{\partial^2\phi(\textbf{\em x})}{\partial x_i^2}dt.
\end{align}
However, using Eq.~(\ref{stationary}) we may represent this as
\begin{align}
  &\langle d\Delta S_1\rangle=\nonumber\\
&\sum_i\int d\textbf{\em x}\;\frac{D_i(\textbf{\em x})}{p(\textbf{\em x})}\left(\frac{\partial p(\textbf{\em x})}{\partial x_i^2}\right)^2dt+p(\textbf{\em x})D_i(\textbf{\em x})\left(\frac{\partial \phi(\textbf{\em x})}{\partial x_i^2}\right)^2dt\nonumber\\
&-2p(\textbf{\em x})\frac{\partial D_i(\textbf{\em x})}{\partial x_i}\frac{\partial \phi(\textbf{\em x})}{\partial x_i}-2p(\textbf{\em x})D_i(\textbf{\em x})\frac{\partial^2\phi(\textbf{\em x})}{\partial x_i^2}dt\nonumber\\
&+p(\textbf{\em x})\frac{\partial^2 D_i(\textbf{\em x})}{\partial x_i^2}dt-\frac{\partial^2(D_i(\textbf{\em x})p(\textbf{\em x}))}{\partial x_i^2}dt-D_i(\textbf{\em x})\frac{\partial^2 p(\textbf{\em x})}{\partial x_i^2}dt.
\end{align}
By further integration by parts, dropping boundary terms and rearranging, this becomes
\begin{align}
&  \langle d\Delta S_1\rangle=\sum_i\int d\textbf{\em x}\;\frac{D_i(\textbf{\em x})}{p(\textbf{\em x})}\left(\frac{\partial p(\textbf{\em x})}{\partial x_i^2}\right)^2dt\nonumber\\
&\qquad+p(\textbf{\em x})D_i(\textbf{\em x})\left(\frac{\partial \phi(\textbf{\em x})}{\partial x_i^2}\right)^2dt+2D_i(\textbf{\em x})\frac{\partial p(\textbf{\em x})}{\partial x_i}\frac{\partial \phi(\textbf{\em x})}{\partial x_i}dt\nonumber\\
\end{align}
which can be written
\begin{align}
\frac{d\langle \Delta S_1\rangle}{dt}&=\frac{d\langle\Delta S_{\rm sys}+\left(\Delta Q_{\rm ex}/k_BT_{\rm env}\right)\rangle}{dt}\nonumber\\
&=\sum_i\int d\textbf{\em x}\frac{ p(\textbf{\em x})}{D_i(\textbf{\em x})}\left(\frac{J_i(\textbf{\em x})}{p(\textbf{\em x})}-\frac{J_i^{\rm st}(\textbf{\em x})}{p^{\rm st}(\textbf{\em x})}\right)^2\nonumber\\
&=\sum_i\int d\textbf{\em x}\frac{ p(\textbf{\em x})}{D_i(\textbf{\em x})}\left(\frac{J_i^{\rm ir}(\textbf{\em x})}{p(\textbf{\em x})}-\frac{J_i^{\rm st,ir}(\textbf{\em x})}{p^{\rm st}(\textbf{\em x})}\right)^2\nonumber\\
\label{S1av}
\end{align}
assuring the positivity of such a contribution. Since it can be written in terms of the total current $J_i$ in this way, it maps precisely onto the non-adiabatic entropy production appearing in \cite{adiabaticnonadiabatic2} and thus can be expressed as
\begin{equation}
  \frac{d\langle \Delta S_1\rangle}{dt}=-\int d{\textbf{\em x}}\;\frac{\partial p(\textbf{\em x})}{\partial t}\ln{\frac{p(\textbf{\em x})}{p^{\rm st}(\textbf{\em x})}}
  \label{S1av2}
\end{equation}
as highlighted by the authors of \cite{adiabaticnonadiabatic2}. We however emphasise that Eq.~(\ref{S1av2}) is to be considered alongside the accompanying SDE in Eq.~(\ref{S1SDE}), from which it has been derived directly, rather than by a division of an observed positive contribution to the mean rate of change of Gibbs entropy into presumed unique transient and stationary terms.\\
\\
We may now by similar means consider an increment in $\Delta S_2$ as follows:
\begin{equation}
  d\Delta S_2=\ln{\frac{p(\textbf{\em x}',t\!+\!dt|\textbf{\em x},t)}{p^{\rm ad}(\boldsymbol{\varepsilon}\textbf{\em x}',t\!+\!dt|\boldsymbol{\varepsilon}\textbf{\em x},t)}}.
  \label{dS2}
\end{equation}
In this case the construction of the denominator follows slightly different rules since, unlike $\Delta S_{\rm tot}$ and $\Delta S_1$, the alternative path, $\vec{\textbf{\em x}}^{\rm T}$, is based on a time reversal of the coordinates, but otherwise follows the sequence of the forward path. As such $b$ behaves in the same manner as $a$ rendering $r_i'=bx_i'+(1-b)x_i$, $f(\textbf{\em r}')dx_i=f(\textbf{\em x})dx_i+2bD(\textbf{\em x})\partial f(\textbf{\em x})/\partial x_i dt$, as detailed in appendix \ref{appendixA}. In this case the appropriate choice for the equivalence of evaluation points $\textbf{\em r}'=\textbf{\em r}$ is $a=b$. For continuity, we may once again choose $a=b=1/2$ with Stratonovich multiplication rules: we represent the transition probability appearing in the numerator through Eq.~(\ref{propagator}), and the denominator by a similar means using the drift term given in Eq.~(\ref{adjointA}) and the path choice $\textbf{\em x}^{\rm T}(t)=\boldsymbol{\varepsilon}\textbf{\em x}(t)$, such that
\begin{widetext}
\begin{center}
\begin{align}
  &p^{\rm ad}(\boldsymbol{\varepsilon}\textbf{\em x}',t+dt|\boldsymbol{\varepsilon}\textbf{\em x},t)=\prod_i\sqrt{\frac{1}{4\pi D_{i}(\boldsymbol{\varepsilon}\textbf{\em r})dt}}\nonumber\\
&\times\exp\left[-\frac{\left(\varepsilon_idx_i+(A_i(\boldsymbol{\varepsilon}\textbf{\em r})-2(\partial D_i(\boldsymbol{\varepsilon}\textbf{\em r})/\partial (\varepsilon_ir_i))+2D_i(\boldsymbol{\varepsilon}\textbf{\em r})(\partial \phi(\boldsymbol{\varepsilon}\textbf{\em r}) /\partial (\varepsilon_ir_i)))dt+(\partial D_{i}(\boldsymbol{\varepsilon}\textbf{\em r})/\partial (\varepsilon_i r_i))dt\right)^2}{4D_{i}(\boldsymbol{\varepsilon}\textbf{\em r})dt}\right.\nonumber\\
      &\left.\qquad+\frac{dt}{2}\frac{\partial }{\partial \varepsilon_ir_i}\left(A_{i}(\boldsymbol{\varepsilon}\textbf{\em r})-2\frac{\partial D_i(\boldsymbol{\varepsilon}\textbf{\em r})}{\partial (\varepsilon_ir_i)}+2D_i(\boldsymbol{\varepsilon}\textbf{\em r})\frac{\partial \phi(\boldsymbol{\varepsilon}\textbf{\em r}) }{\partial (\varepsilon_ir_i)}\right)+\frac{dt}{4}\frac{\partial^2 D_{i}(\boldsymbol{\varepsilon}\textbf{\em r})}{\partial(\varepsilon_i r_i)^2}\right].
\end{align}
\end{center}
We can utilise the usual transformation rules and assumptions for $A^{\rm ir}$, $A^{\rm rev}$ and $D_i$ and express $\partial \phi(\boldsymbol{\varepsilon}\textbf{\em r}) /\partial (\varepsilon_ir_i)=\varepsilon_i\partial \phi(\boldsymbol{\varepsilon}\textbf{\em r}) /\partial r_i=\phi_i'(\boldsymbol{\varepsilon}\textbf{\em r})$ (along with $\partial^2 \phi(\boldsymbol{\varepsilon}\textbf{\em r}) /\partial (\varepsilon_ir_i)^2 =\phi''_i(\boldsymbol{\varepsilon}\textbf{\em x})$) such that we can write the propagator as
\begin{center}
\begin{align}
  &p^{\rm ad}(\boldsymbol{\varepsilon}\textbf{\em x}',t+dt|\boldsymbol{\varepsilon}\textbf{\em x},t)=\prod_i\sqrt{\frac{1}{4\pi D_{i}(\textbf{\em r})dt}}\nonumber\\
&\times\exp\left[-\frac{\left(dx_i+(A^{\rm ir}_i(\textbf{\em r})-A^{\rm rev}_i(\textbf{\em r})-2(\partial D_i(\textbf{\em r})/\partial r_i)+2\varepsilon_iD_i(\textbf{\em r}) \phi_i'(\boldsymbol{\varepsilon}\textbf{\em r})     )dt+(\partial D_{i}(\textbf{\em r})/\partial r_i)dt\right)^2}{4D_{i}(\textbf{\em r})dt}\right.\nonumber\\
      &\left.\qquad+\frac{dt}{2}\frac{\partial }{\partial r_i}\left(A^{\rm ir}_i(\textbf{\em r})-A^{\rm rev}_i(\textbf{\em r})-2\frac{\partial D_i(\textbf{\em r})}{\partial r_i}+2\varepsilon_iD_i(\textbf{\em r})\phi_i'(\boldsymbol{\varepsilon}\textbf{\em r})\right)+\frac{dt}{4}\frac{\partial^2 D_{i}(\textbf{\em r})}{\partial r_i^2}\right].
\end{align}
\end{center}
\end{widetext}
Constructing the ratio in Eq.~(\ref{dS2}) we find
\begin{align}
&d\Delta S_2=\sum_i-\frac{A_i^{\rm ir}(\textbf{\em x})A_i^{\rm rev}(\textbf{\em x})}{D_i(\textbf{\em x})}dt+\frac{A_i^{\rm ir}(\textbf{\em x})}{D_i(\textbf{\em x})}\circ dx_i\nonumber\\
&-\frac{\partial A_i^{\rm ir}(\textbf{\em x})}{\partial x_i}dt+D_i(\textbf{\em x})(\phi_i'(\boldsymbol{\varepsilon}\textbf{\em x}))^2dt-2\varepsilon_i\frac{\partial D_i(\textbf{\em x})}{\partial x_i}\phi_i'(\boldsymbol{\varepsilon}\textbf{\em x})dt\nonumber\\
&+\varepsilon_i(A_i^{\rm ir}(\textbf{\em x})-A_i^{\rm rev}(\textbf{\em x}))\phi_i'(\boldsymbol{\varepsilon}\textbf{\em x})dt+\frac{A_i^{\rm rev}(\textbf{\em x})}{D_i(\textbf{\em x})}{\frac{\partial D_i(\textbf{\em x})}{\partial x_i}}dt\nonumber\\
&+\varepsilon_i\phi_i'(\boldsymbol{\varepsilon}\textbf{\em x})\circ dx_i-\frac{1}{D_i(\textbf{\em x})}\frac{\partial D_i(\textbf{\em x})}{\partial x}\circ dx_i\nonumber\\
&+\frac{\partial^2D_i(\textbf{\em x})}{\partial x_i^2}dt-D_i(\textbf{\em x})\phi_i''(\boldsymbol{\varepsilon}\textbf{\em x})dt
\end{align}
which is the same as
\begin{align}
d\Delta S_2&=d\left(\Delta Q_{\rm hk,G}/k_BT_{\rm env}\right)\nonumber\\
&=\sum_i-\frac{A_i^{\rm ir}(\textbf{\em x})A_i^{\rm rev}(\textbf{\em x})}{D_i(\textbf{\em x})}dt+\frac{A_i^{\rm ir}(\textbf{\em x})}{D_i(\textbf{\em x})}dx_i\nonumber\\
&+\varepsilon_i\phi_i'(\boldsymbol{\varepsilon}\textbf{\em x})dx_i-\frac{1}{D_i(\textbf{\em x})}\frac{\partial D_i(\textbf{\em x})}{\partial x} dx_i\nonumber\\
&+\frac{1}{D_i(\textbf{\em x})}\left(\frac{\partial D_i(\textbf{\em x})}{\partial x_i}\right)^2dt+D_i(\textbf{\em x})(\phi_i'(\boldsymbol{\varepsilon}\textbf{\em x}))^2dt\nonumber\\
&\!\!\!\!\!\!\!\!\!-2\varepsilon_i\phi_i'(\boldsymbol{\varepsilon}\textbf{\em x})\frac{\partial D_i(\textbf{\em x})}{\partial x}dt+\varepsilon_i(A_i^{\rm ir}(\textbf{\em x})-A_i^{\rm rev}(\textbf{\em x}))\phi_i'(\boldsymbol{\varepsilon}\textbf{\em x})dt\nonumber\\
&-\frac{(A_i^{\rm ir}(\textbf{\em x})-A_i^{\rm rev}(\textbf{\em x}))}{D_i(\textbf{\em x})}{\frac{\partial D_i(\textbf{\em x})}{\partial x_i}}dt.
\label{S2SDE}
\end{align}
By employing the averaging procedure we find
\begin{align}
&\frac{d  \langle \Delta S_2\rangle}{dt}=\nonumber\\
&\sum_i\int d\textbf{\em x}\; \frac{p(\textbf{\em x})}{D_i(\textbf{\em x})}\left(A_i^{\rm ir}(\textbf{\em x})-\frac{\partial D_i(\textbf{\em x})}{\partial x_i}+\varepsilon_iD_i(\textbf{\em x})\phi_i'(\boldsymbol{\varepsilon}\textbf{\em x})\right)^2
\end{align}
which may be written
\begin{align}
  \frac{d  \langle \Delta S_2\rangle}{dt}&=\frac{d\langle\Delta Q_{\rm hk,G}/k_BT_{\rm env}\rangle}{dt}\nonumber\\
&=\sum_i\int d\textbf{\em x}\; \frac{p(\textbf{\em x})}{D_i(\textbf{\em x})}\left(\frac{J_i^{\rm ir,st}(\boldsymbol{\varepsilon}\textbf{\em x})}{p^{\rm st}(\boldsymbol{\varepsilon}\textbf{\em x})}\right)^2.
 \label{S2av}
\end{align}
Such a form illustrates the positivity requirement of $\Delta S_2$ in the mean, resulting from its adherence to an IFT, and again Eq.~(\ref{S2av}) is to be considered alongside the complementary SDE in Eq.~(\ref{S2SDE}). Since it is based on an integral over the stationary irreversible flux, $d\langle\Delta S_2\rangle/dt$ describes a contribution to entropy production which arises from an absence of detailed balance and is non-zero both in and out of stationarity. This quantity is to be contrasted with the adiabatic entropy production in \cite{adiabaticnonadiabatic2} which we may now consider to be a special case when there are only even variables in the dynamics. We point out again the importance of the direct derivation of this result from the SDE in this formalism, as opposed to a division of the irreversible flux into terms with structure based solely on $p^{\rm st}(\textbf{\em x})$, which would not have obviously led to the above expression.\\
\\
 We note that the integral in Eq.~(\ref{S2av}) must reduce to the total entropy production, and thus an integral over the stationary irreversible flux (i.e. without the $\boldsymbol{\varepsilon}$ factors inside the squared term in the integrand), in the stationary state, but there are other circumstances when this correspondence applies more generally. A first case is when the irreversible stationary flux is proportional to the stationary distribution, which would be the case for a non-equilibrium constraint that is independent of the phase space variables, as illustrated later in example II, and a second case is when the total flux in each coordinate is everywhere zero ($J_i(\textbf{\em x})=0$), such as for independent variables, $x_i$, defined on regions with natural or reflecting boundaries.\\
\\
To complete the description of all three contributions to entropy production we now consider an increment in $\Delta S_3$.
By using the definition in Eq.~(\ref{S3orig})
\begin{equation}
d\Delta S_{\rm 3}=\ln{\frac{p^{\rm ad}(\boldsymbol{\varepsilon}\textbf{\em x}',t+dt|\boldsymbol{\varepsilon}\textbf{\em x},t)p^{\rm ad}(\textbf{\em x},t+dt|\textbf{\em x}',t)}{p(\textbf{\em x}',t+dt|\textbf{\em x},t)p(\boldsymbol{\varepsilon}\textbf{\em x},t+dt|\boldsymbol{\varepsilon}\textbf{\em x}',t)}}
\end{equation}
together with the previously used propagators, and employing the stationarity condition evaluated at $\boldsymbol{\varepsilon}\textbf{\em x}$:
\begin{align}
&\nabla\cdot J^{\rm st}(\boldsymbol{\varepsilon}\textbf{\em x})=0\nonumber\\
 &=\sum_i\left(-(A^{\rm ir}_i(\boldsymbol{\varepsilon}\textbf{\em x})+A^{\rm rev}_i(\boldsymbol{\varepsilon}\textbf{\em x}))\phi_i'(\boldsymbol{\varepsilon}\textbf{\em x})+\frac{\partial A^{\rm ir}_i(\boldsymbol{\varepsilon}\textbf{\em x})}{\partial (\varepsilon_ix_i)}\right.\nonumber\\
&\left.+\frac{\partial A^{\rm rev}_i(\boldsymbol{\varepsilon}\textbf{\em x})}{\partial (\varepsilon_ix_i)}-D_i(\boldsymbol{\varepsilon}\textbf{\em x})(\phi_i'(\boldsymbol{\varepsilon}\textbf{\em x}))^2\right.\nonumber\\
&\left.-\frac{\partial^2D_i(\boldsymbol{\varepsilon}\textbf{\em x})}{\partial (\varepsilon_ix_i)^2}+D_i(\boldsymbol{\varepsilon}\textbf{\em x})\phi_i''(\boldsymbol{\varepsilon}\textbf{\em x})+2\frac{\partial D_i(\boldsymbol{\varepsilon}\textbf{\em x})}{\partial (\varepsilon_ix_i)}\phi_i'(\boldsymbol{\varepsilon}\textbf{\em x})\right)e^{-\phi(\boldsymbol{\varepsilon}\textbf{\em x})}
\end{align}
we find
\begin{align}
  d\Delta S_3 &=d\left(\Delta Q_{\rm hk,T}/k_BT_{\rm env}\right)\nonumber\\
&=\sum_i\phi_i'(\textbf{\em x})\circ dx_i-\varepsilon_i\phi_i'(\boldsymbol{\varepsilon}\textbf{\em x})\circ dx_i\nonumber\\
&=\sum_i\ln{\frac{\exp{[-\phi(\textbf{\em x})]}}{\exp{[-\phi(\textbf{\em x}')]}}\frac{\exp{[-\phi(\boldsymbol{\varepsilon}\textbf{\em x}')]}}{\exp{[-\phi(\boldsymbol{\varepsilon}\textbf{\em x})]}}}.
\label{S3SDE}
\end{align}
which maps onto the same quantity derived from a master equation approach \cite{prl}. We can then construct the average contribution by converting to Ito form and performing the path integral such that
\begin{align}
  \langle d\Delta S_3\rangle&=\sum_i\int d\textbf{\em x}\;p(\textbf{\em x})A_i(\textbf{\em x})(\phi_i'(\textbf{\em x})-\varepsilon_i\phi_i'(\boldsymbol{\varepsilon}\textbf{\em x}))dt\nonumber\\
  &\quad+p(\textbf{\em x})D_i(\textbf{\em x})(\phi_i''(\textbf{\em x})-\phi_i''(\boldsymbol{\varepsilon}\textbf{\em x}))dt
\end{align}
and proceed to manipulate by integrating by parts, assuming the probability density and current vanish or cancel at boundaries, such that
\begin{align}
\langle d\Delta S_3\rangle&=\sum_i\int d\textbf{\em x}\;p(\textbf{\em x})A_i(\textbf{\em x})(\phi_i'(\textbf{\em x})-\varepsilon_i\phi_i'(\boldsymbol{\varepsilon}\textbf{\em x}))dt\nonumber\\
   &\quad-\int d\textbf{\em x}\;\frac{\partial}{\partial x_i}\left(p(\textbf{\em x})D_i(\textbf{\em x})\right)(\phi_i'(\textbf{\em x})-\varepsilon_i\phi_i'(\boldsymbol{\varepsilon}\textbf{\em x}))dt\nonumber\\
&=\sum_i\int d\textbf{\em x}\;\left(\phi_i'(\textbf{\em x})-\varepsilon_i\phi_i'(\boldsymbol{\varepsilon}\textbf{\em x})\right)\nonumber\\
&\quad\times\left(A_i(\textbf{\em x})p(\textbf{\em x})-\frac{\partial}{\partial x_i}\left(p(\textbf{\em x})D_i(\textbf{\em x})\right)\right)dt\nonumber\\
&=\sum_i\int d\textbf{\em x}\;\left(\phi_i'(\textbf{\em x})-\varepsilon_i\phi_i'(\boldsymbol{\varepsilon}\textbf{\em x})\right)J_i(\textbf{\em x})dt\nonumber\\
&\quad-\int d\textbf{\em x}\;\left(\phi(\textbf{\em x})-\phi(\boldsymbol{\varepsilon}\textbf{\em x})\right)\frac{\partial J_i(\textbf{\em x})}{\partial x_i}dt\nonumber\\
&=-\int d{\textbf{\em x}}\; \left(\phi(\textbf{\em x})-\phi(\boldsymbol{\varepsilon}\textbf{\em x})\right)\left(\sum_i \frac{\partial J_i(\textbf{\em x})}{\partial x_i}\right)dt\nonumber\\
&=-\int d{\textbf{\em x}}\; \left(\phi(\textbf{\em x})-\phi(\boldsymbol{\varepsilon}\textbf{\em x})\right)\left(\nabla\cdot J(\textbf{\em x})\right)dt.
\end{align}
By substituting the original Fokker-Planck equation we may also write this as
\begin{align}
\frac{d\langle \Delta S_3\rangle}{dt}&=\frac{d\langle\Delta Q_{\rm hk,T}/k_BT_{\rm env}\rangle}{dt}\nonumber\\
&=\int d{\textbf{\em x}}\; \frac{\partial p(\textbf{\em x})}{\partial t}\left(\phi(\textbf{\em x})-\phi(\boldsymbol{\varepsilon}\textbf{\em x})\right)\nonumber\\
&=-\int d{\textbf{\em x}}\; \frac{\partial p(\textbf{\em x})}{\partial t}\ln{\frac{p^{\rm st}(\textbf{\em x})}{p^{\rm st}(\boldsymbol{\varepsilon}\textbf{\em x})}}.
\label{S3av}
\end{align}
This has a form similar to Eq.~(\ref{S1av2}) and is clearly a contribution to the mean total entropy production rate that behaves transiently in a manner similar to $\Delta S_1$. The quantity $\Delta S_1$ appears in the Hatano-Sasa relation which describes the entropy production associated with a transition between different stationary states. However, in light of Eq.~(\ref{S3av}) we suggest that $\Delta S_1$, and thus the Hatano-Sasa relation and non-adiabatic entropy production, do not represent the entire entropy production associated with transitions between stationary states (or more generally relaxation) since, in the mean, we can construct a new quantity which comprises all contributions which are non-zero only during relaxation, by combining Eqs.~(\ref{S1av2}) and (\ref{S3av}) giving
\begin{align}
  \frac{d\langle \Delta S_1+\Delta S_3\rangle}{dt}&=\frac{d\langle\Delta S_{\rm sys}+\left(\Delta Q_{\rm ex}+\Delta Q_{\rm hk,T}\right)/k_BT_{\rm env}\rangle}{dt}\nonumber\\
&=-\int d{\textbf{\em x}}\; \frac{\partial p(\textbf{\em x})}{\partial t}\ln{\frac{p(\textbf{\em x})}{p^{\rm st}(\boldsymbol{\varepsilon}\textbf{\em x})}}.
\label{S1S3}
\end{align}
This describes a contribution to the mean entropy production rate which occurs when the system is out of stationarity, but it does not obey an IFT and thus has no guarantee of positivity. \\\\
Our central results therefore are expressions for three contributions to entropy production for arbitrary systems with odd and even dynamical variables evolving according to Ito SDEs with multiplicative noise. These expressions apply to individual trajectories (Eqs.~(\ref{StotSDE}), (\ref{S1SDE}), (\ref{S2SDE}) and (\ref{S3SDE})) and in the mean (Eqs.~(\ref{StotAV}), (\ref{S1av}), (\ref{S2av}) and (\ref{S3av})). Such a demonstration shows the additional complexity introduced by the inclusion of odd variables if one insists on considering entropy production to be due to relaxation or to non-equilibrium constraints with particular reference to Eq.~(\ref{S1S3}). One may think of $\langle\Delta S_1+\Delta S_3\rangle$ as describing a transient contribution to entropy production in the same manner as $\langle\Delta S_1\rangle$, but with the further specification of the nature of the coordinates: the entropy production depends on whether the variables being described are odd or even. The additional complexity of $\Delta S_3$ arises because Eq.~(\ref{S1S3}) can only differ from Eq.~(\ref{S1av2}) when the stationary state is out of equilibrium, such that $p^{\rm st}(\textbf{\em x})\neq p^{\rm st}(\boldsymbol{\varepsilon}\textbf{\em x})$.\\
\\
We of course expect and require that the contributions detailed here are related such that
\begin{equation}
  \frac{d\langle \Delta S_{\rm tot}\rangle}{dt}=\frac{d\langle \Delta S_{1}\rangle}{dt}+\frac{d\langle \Delta S_{2}\rangle}{dt}+\frac{d\langle \Delta S_{3}\rangle}{dt}
\end{equation}
yet their forms derived above do not obviously lend themselves to such a demonstration immediately. For completeness this is shown in appendix \ref{appendixB}.
\section{Example I: Stationary Heat Transport}
\begin{figure}
\includegraphics[width=\columnwidth,clip]{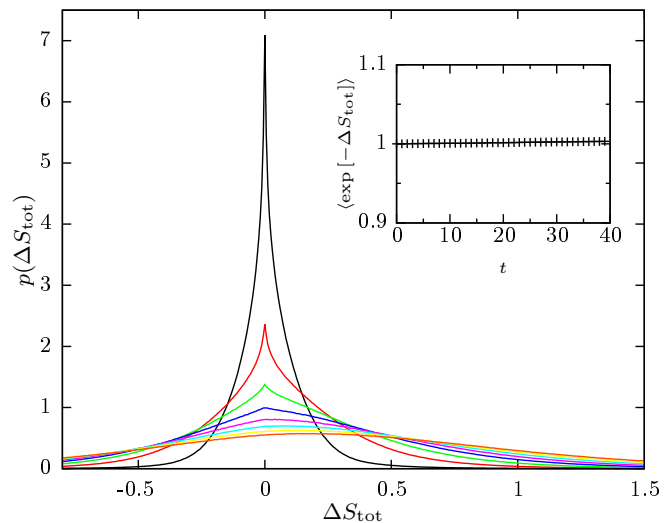}
\caption{\label{fig:Stot} Distributions of dimensionless total entropy production $\Delta S_{\rm tot}$, for example I, for $\gamma=10$ together with a demonstration of adherence to an IFT. Distributions shown are for process intervals from $t=2$ (narrowest) to $t=44$ (widest) in steps of $6$ units.}
\end{figure}

\begin{figure}
\includegraphics[width=\columnwidth,clip]{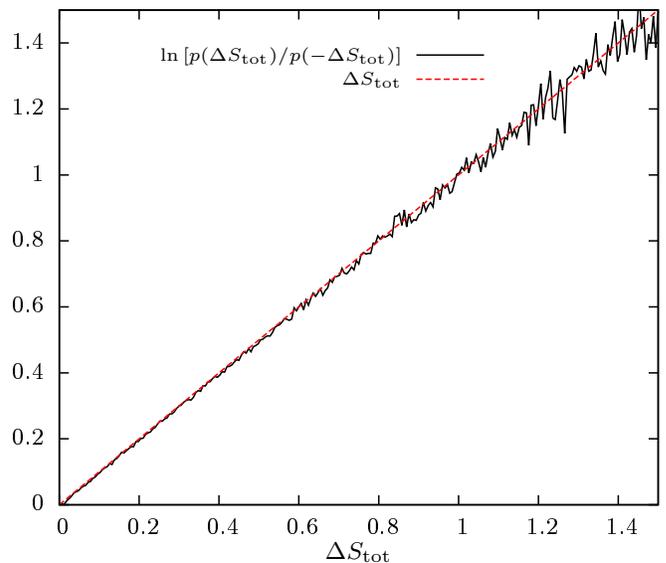}
\caption{\label{fig:DFT} Verification of a detailed fluctuation theorem for example I using data from simulation for $\gamma=10$ at time $t=8$.}
\end{figure}

\begin{figure}
\includegraphics[width=\columnwidth,clip]{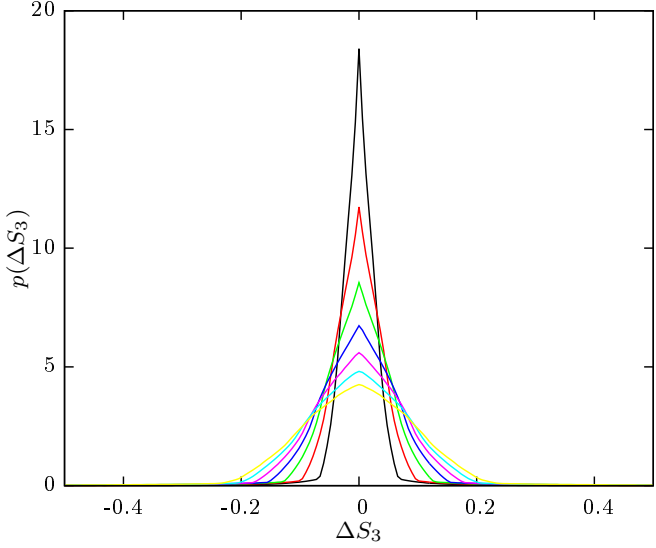}
\caption{\label{fig:S3} Distributions of $\Delta S_3$ for example I evaluated at $t=8$ for a range of $\gamma$ from $\gamma^{-1}=0.1$ (narrowest) to $\gamma^{-1}=0.4$ (widest).}
\end{figure}

We provide as a first example of usage of the above formalism a physical situation which necessitates the use of odd variables in order to describe entropy production adequately: heat transport due to diffusion in one spatial dimension in the presence of a spatially dependent temperature field. Mathematically this system may be modelled without odd (velocity) variables by employing the overdamped limit and constructing a multiplicative SDE and Fokker-Planck equation of the form
\begin{equation}
  dx=\frac{F(x)}{m\gamma}dt+\sqrt{\frac{2k_BT(x)}{m\gamma}}dW
\end{equation}
and
\begin{equation}
  \frac{\partial p(x,t)}{\partial t}=-\frac{\partial}{\partial x}\left(\frac{F(x)p(x,t)}{m\gamma}\right)+\frac{\partial^2}{\partial x^2}\left(\frac{k_BT(x)p(x,t)}{m\gamma}\right)
\end{equation}
where $m$ is the particle mass, $\gamma$ the damping coefficent and $F(x)$ the force operating on the particle. We note the Ito form of both (for a discussion of the resolution of the Ito-Stratonovich dilemma in this case see, for example, \cite{itostrat1,itostrat2,stol}). This Fokker-Planck equation has a stationary distribution
\begin{equation}
  p^{\rm st}(x)=\frac{\mathcal{N}m\gamma}{k_BT(x)}\exp{\left[\int_0^xdx'\;\frac{F(x')}{k_BT(x')}\right]}
\end{equation}
where $\mathcal{N}$ is a normalisation constant. We can quite readily identify the terms $A^{\rm ir}_x=F(x)/m\gamma$, $A^{\rm rev}_x=0$ and $D_x(x)=k_BT(x)/m\gamma$. However, when we come to construct the dimensionless entropy production in the stationary state from Eq.~(\ref{StotSDE}) as
\begin{align}
  &d\Delta S_{\rm tot}=\nonumber\\
&\frac{A^{\rm ir}_{x}(x)}{D_{x}(x)}\circ dx-\frac{1}{D_{x}(x)}\frac{\partial D_{x}(x)}{\partial x}\circ dx-\frac{1}{p^{\rm st}(x)}\frac{\partial p^{\rm st}(x)}{\partial x}\circ dx\nonumber\\
&=\left[\frac{F(x)}{k_BT(x)}-\frac{1}{T(x)}\frac{\partial T(x)}{\partial x}\right.\nonumber\\
&\left.-\frac{1}{p^{\rm st}(x)}\left(-\frac{1}{T(x)}\frac{\partial T(x)}{\partial x}p^{\rm st}(x)+\frac{F(x)}{k_BT(x)}p^{\rm st}(x)\right)\right]\circ dx\nonumber\\
&=0
\end{align}
we find that there is zero entropy production for all trajectories. This may be understood either physically by recognising that in the overdamped limit one demands that the velocity distribution relaxes instantaneously thereby preventing any heat transfer due to temperature inhomogeneities, or geometrically by recognising the impossibility of having stationary flow, and thus entropy production, for a system in one dimension with natural boundaries.\\
\\
To provide a satisfactory representation and to understand the entropy production in such a system we need to consider the more realistic underdamped dynamics in full phase space where we retain both position and velocity coordinates, $x$ and $v$, which are even and odd under time reversal, respectively. The SDEs and Fokker-Planck equation are now given as
\begin{align}
  dx&=vdt\nonumber\\
  dv&=-\gamma vdt+\frac{F(x)}{m}dt+\sqrt{\frac{2k_BT(x)\gamma}{m}}dW
\end{align}
and
\begin{align}
& \frac{\partial p(x,v,t)}{\partial t}=-v\frac{\partial p(x,v,t)}{\partial x}\nonumber\\
  &-\frac{\partial}{\partial v} \left(\left(\frac{F(x)}{m}-\gamma v\right)p(x,v,t)\right)+\frac{k_BT(x)\gamma}{m}\frac{\partial^2 p(x,v,t)}{\partial v^2}.
\end{align}
We may then identify the terms $A_x^{\rm ir}=0$, $A_x^{\rm rev}=v$, $A_v^{\rm ir}=-\gamma v$, $A_v^{\rm rev}=F(x)/m$, $D_x=0$ and $D_v=k_BT(x)\gamma/m$. By Eq.~(\ref{StotSDE}) the dimensionless entropy production is
\begin{align}
  &d\Delta S_{\rm tot}\nonumber\\
&=-d(\ln p(x,v,t))-\frac{mv}{k_BT(x)}\circ dv+\frac{Fv}{k_BT(x)}dt\nonumber\\
 &=-d(\ln p(x,v,t))-\frac{1}{k_BT(x)}d\left(\frac{mv^2}{2}\right)+\frac{F}{k_BT(x)}dx
\label{SDEentv}
\end{align}
using $v\circ dv=(1/2)(v'+v)(v'-v)$ and $vdt=dx$, and noting that $x$ is now deterministic, meaning the integration rules are irrelevant. The second and third terms correctly reproduce the form of the change in medium entropy as heat transfer to the environment, equal to negative heat transfer to the particle (in agreement with the result found in stochastic energetics \cite{sekimotobook}), divided by the instantaneous temperature, and do so only by virtue of the consideration of odd and even variables.\\
\\
We can use this SDE to produce distributions of entropy production and verify relevant fluctuation theorems. To do so, however, requires knowledge of the solution to the Fokker-Planck equation, for which there is no simple analytical form. To proceed we restrict ourselves to the stationary state and utilise the expansion found in \cite{stol} and \cite{Widder} which expresses the stationary solution as a series expansion about the overdamped distribution:
\begin{align}
  p^{\rm st,over}(x,v)&=\frac{\mathcal{N}m}{k_BT(x)}\exp{\left[\int_0^x dx'\;\frac{F(x')}{k_BT(x')}\right]}\nonumber\\
&\quad\times\sqrt{\frac{m}{2\pi k_BT(x)}}\exp{\left[-\frac{mv^2}{2k_BT(x)}\right]},
\label{stolapprox}
\end{align}
where $\mathcal{N}$ is determined by normalisation, such that
\begin{equation}
  p^{\rm st}(x,v)=p^{\rm st,over}(x,v)+\sum_{i=1}^{\infty}(1/\gamma)^{i}p_i(x,v)
\label{expan}
\end{equation}
$p_i(x,v)$ has a general form
\begin{align}
  &p_i(x,v)=\nonumber\\
&\sum_{k=a_i}^{k=b_i}\frac{c_{i,k}(x)H_k(v\sqrt{m/k_BT(x)})}{\sqrt{2\pi k_BT(x)/m}}\exp{\left[-\frac{mv^2}{2k_BT(x)}\right]},
\end{align}
where constants $a_i$, $b_i$ and functions $c_{i,k}(x)$ are found by an iterative procedure, and $H_k(y)$ are Hermite polynomials defined as
\begin{equation}
  H_k(y)=(-1)^ke^{\frac{y^2}{2}}\frac{d^k}{dy^k}e^{\frac{-y^2}{2}}.
\end{equation}
Whilst the expansion has the formal deficiency that the expansion parameter is not unitless it suffices for a theoretical illustration where we can consider it in a limit where it is appropriate. We consider units $k_B=T=m=1$, a harmonic confining potential such that $F(x)=-x$, a temperature profile
\begin{equation}
  T(x)=1+\frac{1}{2}\tanh(x),
\end{equation}
and approximate the stationary distribution by considering the expansion in Eq.~(\ref{expan}) to fourth order in $\gamma^{-1}$, applying the formalism numerically.\\
\\
We first demonstrate that this approach yields a result which maps onto the expected phenomenological expression for dimensionless internal entropy generation \cite{chaikin}
\begin{equation}
\frac{d\langle \Delta S_{\rm tot}\rangle}{dt} =\int_{-\infty}^{+\infty}dx\;J_Q(x)\frac{\partial}{\partial x}\left(\frac{1}{k_BT(x)}\right)
\label{Sclass}
\end{equation}
where $J_Q(x)$ is the stationary heat current defined as
\begin{equation}
  J_Q(x)
  =\int_{-\infty}^{+\infty} dv\;\frac{1}{2}mv^3p^{\rm st}(x,v).
\end{equation}
\begin{figure}
\includegraphics[width=\columnwidth,clip]{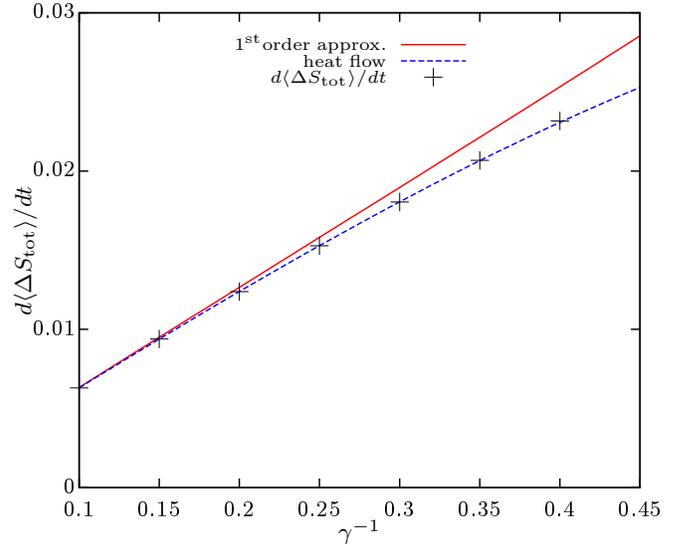}
\caption{\label{fig:Stotav} Mean dimensionless entropy production, for example I, for a range of damping coefficients as predicted by a first order approximation in Eq.~(\ref{1storderent}) (solid line), an integral over the heat current, Eq.~(\ref{Sclass}) (dashed line) and a Monte Carlo average based on the SDE in Eq.~(\ref{SDEentv}) (crosses).}
\end{figure}
Figure \ref{fig:Stotav} shows the dimensionless entropy production obtained by performing the integral in Eq.~(\ref{Sclass})  using a numerically calculated $p^{\rm st}(x,v)$, compared with that obtained by averaging the SDE in Eq.~(\ref{SDEentv}) by Monte Carlo simulation of the underlying particle dynamics, for a range of damping coefficients, alongside a demonstrably positive first order approximation based on the first correction term in Eq.~(\ref{expan}) given by
\begin{equation}
  \frac{d  \langle \Delta S_{\rm tot}\rangle}{dt}\simeq \int^{+\infty}_{-\infty} dx\;\frac{k_Bp^{\rm st}(x)}{2m\gamma T(x)}\left(\frac{\partial T(x)}{\partial x}\right)^2
\label{1storderent}
\end{equation}
where
\begin{equation}
  p^{\rm st}(x)=\int_{-\infty}^{+\infty}dv\; p^{\rm st}(x,v).
\end{equation}
Our formalism for the rate of change of $\langle \Delta S_{\rm tot}\rangle$ agrees with Eq.~(\ref{Sclass}) and both are consistent with Eq.~(\ref{1storderent}) in the $\gamma\to\infty$ limit. We point out that total entropy production decreases as coupling to the environment increases which may seem counter-intuitive, but we emphasise that with increased coupling, despite greater heat transfer to and from the environment, there is highly diminished spatial heat \emph{transport} (the latter being the cause of entropy production) as the system is brought closer to a local equilibrium.\\
\\
 We can use the SDE for entropy production (Eq.~(\ref{SDEentv})) to move beyond a classical description of mean entropy production to one described by Jarzynski, Seifert, Sekimoto and others \cite{sekimoto1,Jarzynski97,seifertprinciples} where we can identify entropy generating and destroying trajectories. We can explicitly calculate the distribution of total entropy production which is shown for $\gamma=10$ in Figure \ref{fig:Stot} for various process intervals, along with a demonstration that it adheres to an IFT throughout. Additionally, since we consider the stationary state we can demonstrate a detailed fluctuation theorem of the form $p(\Delta S_{\rm tot})/p(-\Delta S_{\rm tot})=\exp(\Delta S_{\rm tot})$ \cite{seifertoriginal} as shown in Figure \ref{fig:DFT}.\\
Finally we point out that, being in the stationary state, $d\langle \Delta S_3\rangle/dt=0$, but since it is a non-equilibrium stationary state that is asymmetric in the odd velocity variable we have $\Delta S_3\neq0$ in detail, as is clear in Eq.~(\ref{S3SDE}). We can demonstrate the increasing range of values of $\Delta S_3$ as $\gamma$ is reduced and the system is taken further away from local equilibrium, with its symmetric  velocity distribution, by generating the distribution of $\Delta S_3$ using Eq.~(\ref{S3SDE}) for a given time interval, as shown in Figure \ref{fig:S3}. Such a result highlights the fact that although a non-zero $d\langle \Delta S_3\rangle/dt$ is only possible during relaxation as shown by Eq.~(\ref{S3av}), the specific evolution of $\Delta S_3$ for each trajectory is brought about by non-equilibrium constraints that cause the stationary solution to depart from equilibrium.
\section{Example II: Particle Driven by a Non-conservative Force}
\begin{figure}
\includegraphics[width=\columnwidth,clip]{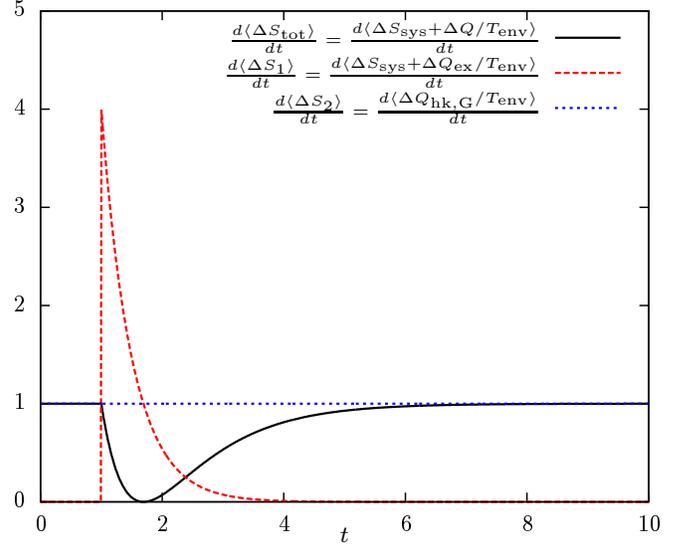}
\caption{\label{fig:one} Positive mean rates of dimensionless entropy change against time for example II, where we consider the transition between stationary states of a driven particle on a ring with $F_0=1$, $F_1=-1$, $t_0=1$ and $k_B=m=\gamma=T=1$.}
\end{figure}

\begin{figure}
\includegraphics[width=\columnwidth,clip]{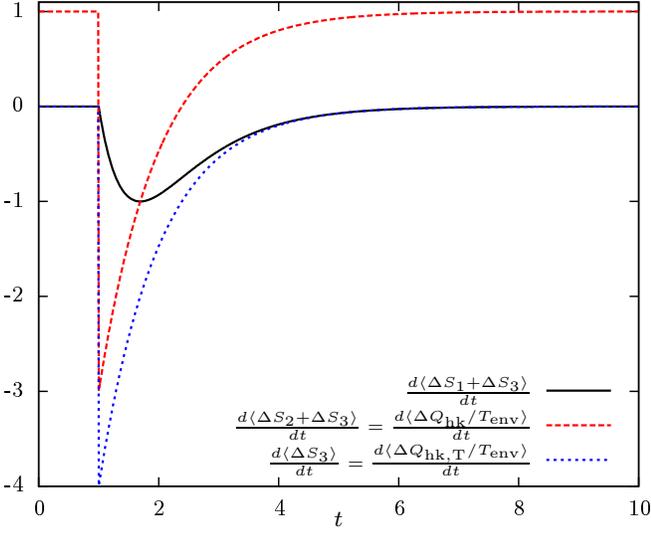}
\caption{\label{fig:two} Unbounded mean rates of dimensionless entropy change for example II, the driven system on a ring with $F_0=1$, $F_1=-1$, $t_0=1$ and $k_B=m=\gamma=T=1$.}
\end{figure}

\begin{figure}
\includegraphics[width=\columnwidth,clip]{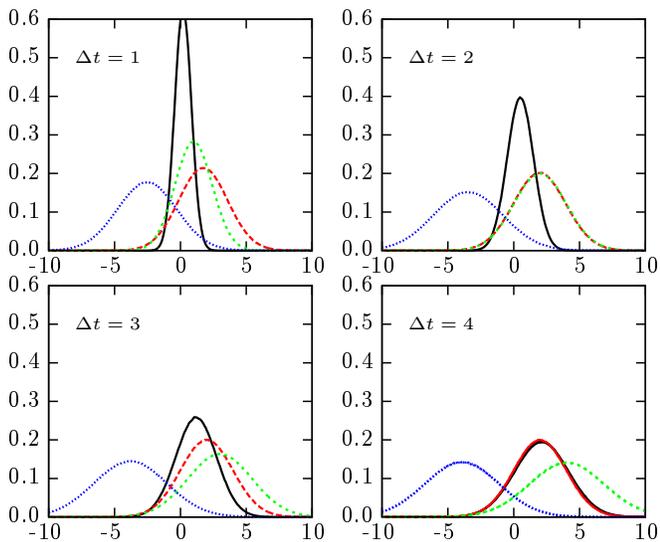}
\caption{\label{fig:s1s2s3} Distributions of entropy productions $\Delta S_{\rm tot}$ (solid line), $\Delta S_1$ (wide dashed line), $\Delta S_2$ (narrow dashed line) and $\Delta S_3$ (dotted line) measured at times $\Delta t=t-t_0=1$, $\Delta t=2$, $\Delta t=3$ and $\Delta t=4$ after the reversal of the force for $F_0=1$, $F_1=-1$, $t_0=1$ and $k_B=m=\gamma=T=1$.}
\end{figure}

\begin{figure}
\includegraphics[width=\columnwidth,clip]{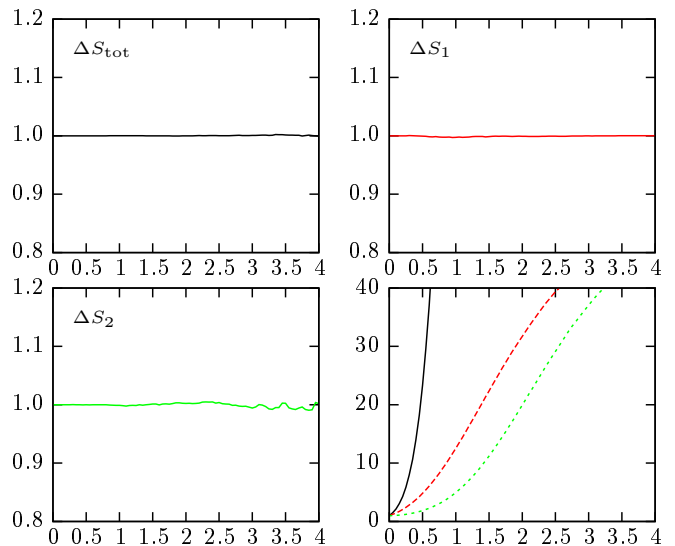}
\caption{\label{fig:s1s2s3IFT} Illustration of adherence to IFTs by consideration of the average $\langle\exp[-\Delta S]\rangle$ against time, $\Delta t=t-t_0$ after the force reversal, for $\Delta S_{\rm tot}$, $\Delta S_1$ and $\Delta S_2$ (indicated) and the failure to adhere to an IFT of $\Delta S_3$ (solid line, fourth subplot), $\Delta S_1+\Delta S_3$ (wide dashed line, fourth subplot) and $\Delta S_2+\Delta S_3$ (narrow dashed line, fourth subplot) for $F_0=1$, $F_1=-1$, $t_0=1$ and $k_B=m=\gamma=T=1$.}
\end{figure}

\begin{figure}
\includegraphics[width=\columnwidth,clip]{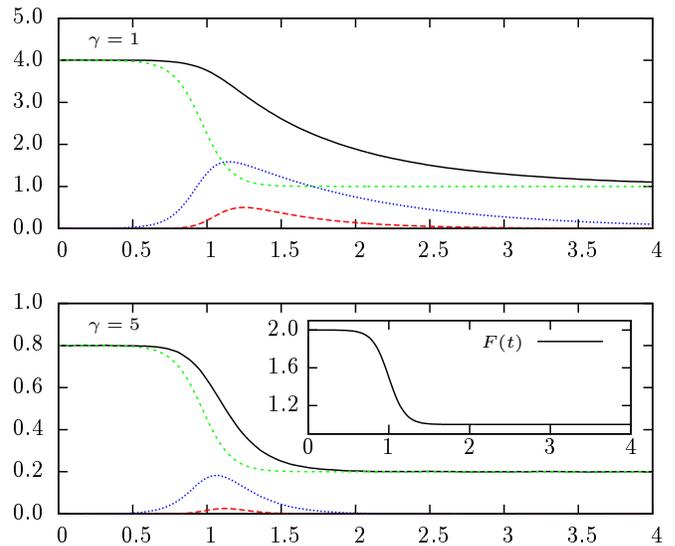}
\caption{\label{fig:tanh} Mean rates of change of $\Delta S_1$ (wide dashed line), $\Delta S_2$  (narrow dashed line), $\Delta S_3$ (dotted line), and their sum $\Delta S_{\rm tot}$ (solid line) for example II with a time dependent force given by Eq.~(\ref{FF}) with units $k_B=m=T=1$, and for $\gamma=1$ (top) and $\gamma=5$ (bottom).}
\end{figure}

Once again utilising the full phase space Langevin description of the dynamics, we consider diffusion of a particle on a ring driven by a spatially independent non-conservative force and spatially independent (additive) noise such that
\begin{align}
  dx&=vdt\nonumber\\
  dv&=-\gamma vdt+\frac{F(t)}{m}dt+\sqrt{\frac{2k_BT\gamma}{m}}dW
\end{align}
thus giving $A_x^{\rm ir}=0$, $A_x^{\rm rev}=v$, $A_v^{\rm ir}=-\gamma v$, $A_v^{\rm rev}=F(t)/m$, $D_x=0$ and $D_v=k_BT\gamma/m$.
For any non-zero value of $F(t)$ there will exist a stationary solution with an asymmetric Gaussian distribution in $v$ and a uniform distribution in $x$ due to the symmetry of the problem. Any relaxation from a given stationary state caused by changes to the non-conservative force will then also result in a uniform distribution in $x$ for all time by the translational symmetry. As such we may proceed by considering the marginalised velocity distribution when starting from a stationary state. Exploiting the fact that the initial Gaussian solution will remain Gaussian for any $F(t)$, we can parameterise a transient solution to the Fokker-Planck equation
\begin{equation}
  p(x,v,t)\propto \sqrt{\frac{m}{2\pi k_BT}}\exp{\left[-\frac{m(v-\langle v\rangle)^2}{2k_BT}\right]}
\end{equation}
with
\begin{equation}
  \frac{d\langle v\rangle}{dt}=\left(\frac{F}{m}-\gamma\langle v\rangle\right)
\end{equation}
such that
\begin{equation}
  \langle v\rangle^{\rm st}=\frac{F}{m\gamma}.
\end{equation}
A scenario where closed form solutions exist for all contributions to entropy production is that of an instantaneous step change in the driving force $F(t)$ so that we have
\begin{equation}
F(t) = \left\{
\begin{array}{rl}
F_0 & \qquad t < t_0,\\
F_1 & \qquad t \geq t_0 ,
\end{array} \right.
\end{equation}
and
\begin{equation}
\langle v\rangle(t) = \left\{
\begin{array}{rl}
{F_0}/{m\gamma} & \quad t < t_0,\\
{\left({F_1+e^{-\gamma (t-t_0)}(F_0-F_1) }\right)}/{m\gamma} & \quad t \geq t_0 .
\end{array} \right.
\end{equation}
Performing the relevant integrals in Eqs.~(\ref{StotAV}), (\ref{S1av}), (\ref{S2av}) and (\ref{S3av}) we then obtain
\begin{equation}
\frac{d\langle\Delta S_{\rm tot}\rangle}{dt} = \left\{
\begin{array}{rl}
{F_0^2}/{m\gamma k_BT} & \quad t < t_0,\\
\left(F_0+F_1(e^{\gamma(t-t_0)}-1)\right)^2 & \quad t \geq t_0, \\
\times{e^{-2\gamma(t-t_0)}}/{m\gamma k_BT}&
\end{array} \right.
\end{equation}
\begin{equation}
\frac{d\langle\Delta S_1\rangle}{dt} = \left\{
\begin{array}{rl}
0 & \quad t < t_0,\\
{e^{-2\gamma(t-t_0)}(F_0-F_1)^2}/{m\gamma k_BT} & \quad t \geq t_0 ,
\end{array} \right.
\end{equation}
\begin{equation}
\frac{d\langle\Delta S_2\rangle}{dt} = \left\{
\begin{array}{rl}
{F_0^2}/{m\gamma k_BT} & \quad t < t_0,\\
{F_1^2}/{m\gamma k_BT} & \quad t \geq t_0 ,
\end{array} \right.
\end{equation}
and
\begin{equation}
\frac{d\langle\Delta S_3\rangle}{dt} = \left\{
\begin{array}{rl}
0 & \quad t < t_0,\\
-{2e^{-\gamma(t-t_0)}F_1(F_1-F_0)}/{m\gamma k_BT} & \quad t \geq t_0 .
\end{array} \right.
\end{equation}
Choosing the specific case of a reversal of the driving force such that it changes from $F_0=1$ to $F_1=-1$ at time $t_0=1$ and employing units $k_B=m=\gamma=T=1$, we can generate the results shown in Figures \ref{fig:one} and \ref{fig:two}.\\
\\
We note first that the mean rates of change of all three contributions $\Delta S_{\rm tot}$, $\Delta S_{1}$ and $\Delta S_{2}$ are positive, reflecting their adherence to an IFT. All three mean rates of change are constant for $t<t_0=1$, are perturbed by the change in direction of the force, and relax back to constant values consistent with the transition between the stationary states. A key feature of this behaviour is that upon perturbation, the total entropy production rate \emph{decreases} which would not emerge using an overdamped description of the dynamics. This feature can be explained by the existence of the $d\langle\Delta S_3\rangle/dt$ contribution to the mean entropy production rate, which may take negative values depending on the relationship between the instantaneous distribution and the stationary distribution. In this specific case, the large negative value for $d\langle \Delta S_3\rangle/dt$ indicates that upon reversal of the force the instantaneous distribution corresponds to particle motion, on average, in a direction counter to that expected to result from the new value of the force. The velocity distribution does relax, of course, to the distribution that corresponds to the new value of the force and so the mean rate of change of $\Delta S_3$ decays away. An important point to draw from Figure~\ref{fig:two} is that $\Delta S_3$, $\Delta S_1+\Delta S_3$ and $\Delta S_2+\Delta S_3$ cannot be expected, in general, to be positive, reflecting that they cannot be expressed in the form of Eq.~(\ref{entform}) and thus do not obey IFTs. This means previous approaches where the entropy production can always be divided into two positive quantities \cite{adiabaticnonadiabatic0,adiabaticnonadiabatic1,adiabaticnonadiabatic2} and the house-keeping heat can be expected to obey an IFT \cite{IFThousekeeping}, do not extend to the systems considered here.\\
\\
We consider this example to be a helpful illustration of how entropy production cannot always be divided into two contributions which derive from relaxation, and an absence of detailed balance owing to a non-equilibrium constraint, respectively. Explicitly, the non-equilibrium constraint here is the constant force which produces entropy in the stationary state by inducing a constant flux around the ring. The mean rate of entropy production in that stationary state is characterised by $d\langle\Delta S_2\rangle/dt$ which remains constant throughout the process owing to the constant magnitude of the force which is applied. However, both $\Delta S_2$ and $\Delta S_3$ are non-zero only in the presence of a non-equilibrium constraint which breaks detailed balance. At the same time the \emph{mean} rate of change of $\Delta S_3$ is non-zero only when the distribution is relaxing to a new stationary solution in the same manner as $\Delta S_1$. Whilst $\Delta S_1$ describes the entropy production that arises from an evolution of the probability distribution of a general set of variables, $\Delta S_3$ expresses what $\Delta S_1$ explicitly leaves out: the additional impact of relaxation on entropy production that relates to the a priori physical specification of the variables as odd or even. Clearly, given that the non-equilibrium constraint is a force of constant magnitude, reflected by the constant $d\langle \Delta S_2\rangle/dt$, it is reasonable to consider the sum of $\Delta S_1$ and $\Delta S_3$ as the contribution that arises due to relaxation to a new stationary state, particularly when the form of its mean rate of change in Figure \ref{fig:two} is contrasted with that of $\Delta S_{\rm tot}$, $\Delta S_1$ and $\Delta S_2$ in Figure \ref{fig:one}. We may make the analysis complete by considering the SDEs for all contributions. The explicit Ito forms of Eqs.~(\ref{StotSDE}), (\ref{S1SDE}), (\ref{S2SDE}) and (\ref{S3SDE}) are given as
\begin{equation}
 d\Delta S_{\rm tot}=-\frac{m}{k_BT}\langle v\rangle dv-\frac{m}{k_BT}\left(v-\langle v\rangle\right)\frac{d\langle v\rangle}{dt}dt+\frac{F(t)}{k_BT}dx
\end{equation}
\begin{equation}
 d\Delta S_{\rm 1}=\frac{1}{k_BT}\left(\frac{F(t)}{\gamma}-m\langle v\rangle\right) dv-\frac{m}{k_BT}\left(v-\langle v\rangle\right)\frac{d\langle v\rangle}{dt}dt
\end{equation}
\begin{equation}
  d\Delta S_{2}=\frac{F(t)}{\gamma k_BT}dv+\frac{F(t)}{k_BT}dx
\end{equation}
\begin{equation}
  d\Delta S_{3}=-\frac{2F(t)}{\gamma k_BT}dv
\end{equation}
and illustrate the behaviour of all the contributions. $d\Delta S_{\rm tot}$ is only zero when $\langle v\rangle=0$, $F=0$ and $d\langle v\rangle/dt=0$ meaning the system is in the equilibrium state. $d\Delta S_1$ is zero whenever $\langle v\rangle=F/m\gamma$ and $d\langle v\rangle/dt=0$ corresponding to any stationary state, equilibrium or otherwise, whilst $d\Delta S_2$ and $d\Delta S_3$ contribute independently of properties of the distribution (namely $\langle v\rangle$), but only when the non-equilibrium constraint is present such that $F(t)\neq0$. $d\Delta S_3$ however, has a mean contribution of zero at stationarity since $\langle dv\rangle=0$ for any stationary state. We can calculate distributions of all the contributions, as measured from the force reversal, numerically using the above SDEs and demonstrate the validity of IFTs, where appropriate, in Figures \ref{fig:s1s2s3} and \ref{fig:s1s2s3IFT}. We observe that all distributions take Gaussian form, to be expected as the model is essentially a recasting of the overdamped dragged oscillator found in \cite{Saha09} where the further, but non-general, detailed fluctuation theorem symmetry $p(\Delta S_{\rm tot})/p(-\Delta S_{\rm tot})=\exp{(\Delta S_{\rm tot})}$  has been noted to hold over finite times \cite{Saha09}, but stressed elsewhere \cite{shargel} to be coincidental. Further insight into this coincidence can be derived from the form of the SDEs which yield Gaussian distributions (for the given initial conditions) since they comprise only drift and additive noise terms (that is, no terms of the form $f(v)dv$). Such properties however, do not distract from the nature of the contributions which can be readily observed: the distributions in $\Delta S_1$ and $\Delta S_3$ develop fastest at first reflecting the initially fast response of the distribution to the change in force. However, distributions for both $\Delta S_2$ and $\Delta S_{\rm tot}$ develop steadily, owing to their contributions being characterised by steady heat dissipation. As such, as time progresses, the distribution of $\Delta S_1$ ceases to develop as the system reaches the new stationary state and the distributions of $\Delta S_2$ and $\Delta S_{\rm tot}$ continue to shift to the right until they eventually dominate. Similarly for $\Delta S_3$, we observe here that the distribution stops evolving despite receiving non-zero contributions.
\\
\\
For completeness we investigate the same model with a less trivial time dependence in the non-conservative force, along with its approach to the overdamped limit where such systems have been considered previously \cite{adiabaticnonadiabatic2,speck2007}. We employ the force protocol
\begin{equation}
F(t) = 1.5-0.5\tanh(-5(t-1))
\label{FF}
\end{equation}
and perform the calculations numerically for two values of damping coefficient, $\gamma=1$ and $\gamma=5$. We point out again that the meaning of ${d\langle \Delta S_2\rangle}/{dt}$ for this system is easily elucidated since the non-equilibrium constraint, $F(t)$, being phase space independent, leads to $J^{\rm ir,st}_{v}\propto p^{\rm st}$ so that
\begin{equation}
  \frac{d\langle \Delta S_2\rangle}{dt}=\frac{d\langle \Delta S_{\rm tot}\rangle^{\rm st}}{dt}=\frac{F(t)^2}{m\gamma k_BT}.
\end{equation}
 The mean contributions for such a protocol for two values of the damping coefficient, again starting from the stationary state, are shown in Figure \ref{fig:tanh}.
Note that in this case the contribution $d\langle\Delta S_3\rangle/dt$ is positive, reflecting that as the non-conservative force decreases, the instantaneous distribution corresponds to a greater average particle flux than would be expected from the instantaneous value of the force, thus producing more entropy in the process of relaxation from one stationary state to the other than would be expected if the relaxation were instantaneous. As $\gamma$ increases, the asymmetry of the stationary state (in velocity) decreases and the contribution from $\Delta S_3$ diminishes. Consequently, the two stationary distributions become increasingly similar, meaning the contribution $\Delta S_1$ also diminishes rendering the total entropy production almost entirely comprised of the contribution from $\Delta S_2$. When the full overdamped limit is taken $\Delta S_2$ is the only contribution and the results map onto those found in \cite{adiabaticnonadiabatic2}.
\section{Discussion and Conclusions}
We have derived SDEs describing the fluctuating evolution of three contributions to entropy production, along with expressions for their mean behaviour, and demonstrated that two of these contributions obey IFTs and thus are rigorously positive in the mean. Furthermore we have demonstrated that whilst these two naturally align themselves with the irreversibility associated with relaxation, and non-equilibrium constraints, respectively, the inclusion of odd dynamical variables can give rise to a third term, which has no bounds on its sign and which cannot be so readily associated with one origin of entropy production or the other. We have sought to make these expressions as general as possible, within reason, with the intention that they may be applied to any system (physical or otherwise) described by stochastic differential equations, providing a framework for the discussion of entropy production, as defined here, within as wide a range of applications as may be relevant. To this end, we have considered a simple heat conduction problem, and after demonstrating that a full phase space representation of the dynamics is crucial to the treatment of its entropy production, we have examined specifically how it may be evaluated. The second example, that of a transition between stationary states of drift-diffusion on a ring, demonstrates the need for a third contribution to entropy production in the analysis, and provides some intuitive understanding of its nature.\\
\\
We suggest that the division of the total entropy production into $\Delta S_1$, $\Delta S_2$ and $\Delta S_3$, as we propose, is always helpful for three main reasons. The first is the ability to identify the physical origins of irreversibility in a process (relaxation and non-equilibrium constraints) and the interplay between them. The second is related to the identification of IFTs with the subsequent positivity requirements and restrictions on the statistics for the two contributions, $\Delta S_1$ and $\Delta S_2$, which unambiguously align themselves with the two causes of irreversibility. And thirdly, the more delicate reason that all three contributions are constructed from total path probability densities with equivalent measures such that they align themselves unambiguously with the same contributions found in master equation approaches which are necessarily formed from path \emph{probabilities} \cite{prl}. This third point may be contrasted with an alternative division of the total entropy production into a system and medium contribution, neither of which can be expressed as ratios of total path probabilities. Being formed from a probability density, the system entropy, as defined in \cite{seifertoriginal} and employed here (Eq.~(\ref{sysmed}) implies $S_{\rm sys}=-\ln{p(\textbf{\em x}(t),t)}$), is strictly not dimensionally correct (even if one argues that the relative entropy change is well defined \cite{seifertprinciples}), but moreover does not share the same form as the system entropy that appears in master equation approaches, which if followed would imply $S_{\rm sys}=-\ln{p(\textbf{\em x}(t),t)d\textbf{\em x}(t)}$. Such ambiguity then also enters into the definition of the medium entropy, but can be avoided altogether by considering the total entropy production in terms of explicit measures of irreversibility such that $\Delta S_{\rm tot}$ is comprised of $\Delta S_1$, $\Delta S_2$ and $\Delta S_3$, all four of which do not suffer from such issues.\\
\\
Generalisations that are immediately obvious beyond the description given here may include correlated stochastic processes for which the method described should prove suitable, and processes where we no longer assume $D(\boldsymbol{\varepsilon}\textbf{\em x})=D(\textbf{\em x})$ for which the generalisation may be more challenging. One might note, however that processes which do not possess such a symmetry are unlikely to be physically meaningful. Furthermore, it would be natural to explore examples involving odd dynamical variables such as angular momentum and magnetic dipole moments, and to include driving by external forces that are themselves odd under time reversal, such as torques and magnetic fields. We expect to find further richness in the phenomenology of entropy production associated with stochastic dynamical behaviour.

\begin{acknowledgments}
RES acknowledges financial support from the UK Engineering and Physical Sciences Research Council.
\end{acknowledgments}
\appendix
\section{The use of short time propagators with multiplicative noise}
\label{appendixA}
Here we consider one of the terms in Eq.~(\ref{lim}) and derive Eq.~(\ref{itomed}). By utilising Eqs.~(\ref{propagator}) and (\ref{backprop}) we can describe an increment in the medium entropy production according to the formalism of Seifert \cite{seifertoriginal} as
\begin{align}
  &d\Delta S_{\rm med}=\nonumber\\
&\sum_i \frac{1}{2}\ln{D_{i}(\textbf{\em r}')}\!-\!\frac{1}{2}\ln{D_{i}(\textbf{\em r})}\!+\!\frac{dx_i^2}{4D_{i}(\textbf{\em r}')dt}\!-\!\frac{dx_i^2}{4D_{i}(\textbf{\em r})dt}\nonumber\\
  &\!+\!\frac{dx_i}{2}\left(\frac{A^{\rm rev}_i(\textbf{\em r})}{D_{i}(\textbf{\em r})}\!+\!\frac{A^{\rm ir}_i(\textbf{\em r})}{D_{i}(\textbf{\em r})}\!+\!\frac{A^{\rm ir}_i(\textbf{\em r}')}{D_{i}(\textbf{\em r}')}\!-\!\frac{A^{\rm rev}_i(r')}{D_{i}(\textbf{\em r}')}\!\right.\nonumber\\
&\left.\qquad\qquad\!-\!2a\frac{1}{D_{i}(\textbf{\em r})}\!\frac{\partial D_{i}(\textbf{\em r})}{\partial r_i}\!-\!2b\frac{1}{D_{i}(\textbf{\em r}')}\!\frac{\partial D_{i}(\textbf{\em r}')}{\partial r'_{i}}\right)\nonumber\\
& \!-\!\frac{dt}{4}\left(\frac{(A^{\rm rev}_i(\textbf{\em r})+A^{\rm ir}_i(\textbf{\em r}))^2}{D_{i}(\textbf{\em r})}-\frac{(A^{\rm rev}_i(\textbf{\em r}')-A^{\rm ir}_i(\textbf{\em r}'))^2}{D_{i}(\textbf{\em r}')}\right)\nonumber\\
&\!-\!adt\left(D_{i}(\textbf{\em r})\frac{\partial }{\partial r_i}\left(\frac{ A^{\rm ir}_i(\textbf{\em r})}{ D_{i}(\textbf{\em r})}\right)+D_{i}(\textbf{\em r})\frac{\partial }{\partial r_i}\left(\frac{ A^{\rm rev}_i(\textbf{\em r})}{ D_{i}(\textbf{\em r})}\right)\right)\nonumber\\
&\!-\!bdt\left(-D_{i}(\textbf{\em r}')\frac{\partial }{\partial r'_{i}}\left(\frac{A^{\rm ir}_i(\textbf{\em r}')}{D_{i}(\textbf{\em r}')}\right)+D_{i}(\textbf{\em r}')\frac{\partial }{\partial r'_{i}}\left(\frac{A^{\rm rev}_i(\textbf{\em r}')}{D_{i}(\textbf{\em r}')}\right)\right)\nonumber\\
&\!+\!a^2dt\left(\frac{\partial^2 D_{i}(\textbf{\em r})}{\partial r_i^2}\!-\!\frac{1}{D_{i}(\textbf{\em r})}\left(\!\frac{\partial D_{i}(\textbf{\em r})}{\partial r_i^2}\!\right)^2\right)\!\nonumber\\
&-\!b^2dt\left(\frac{\partial^2 D_{i}(\textbf{\em r}')}{\partial {r'}_i^2}\!-\!\frac{1}{D_{i}(\textbf{\em r}')}\left(\!\frac{\partial D_{i}(\textbf{\em r}')}{\partial {r'}_i^2}\!\right)^2\right)
\label{ratiofull}
\end{align}
where time dependence in variables $A^{\rm ir}$, $A^{\rm rev}$ and $D_i$ is assumed, but not explicitly written for brevity. We may proceed by understanding that the quantity $dx_i$ is an increment in an underlying SDE, meaning that we must consider all multiplications of the form $f(r)dx$ as infinitesimal stochastic integrals with a summation rule defined by the evaluation point $r$. For example, $r=x$ would imply an Ito integration, $r=(1/2)(x+x')$ would imply Stratonovich and so on. To consolidate the above it is sensible to convert all  multiplications into one type, for which we choose Ito in order to apply the Ito stochastic calculus transparently using the heuristic rules $(dW_i)^2=dt$ and $dW_idW_j=0$, and to drop all terms of order $dt^{3/2}$ and higher. To do so we apply the following reasoning. For a suitably smooth function, $f(\textbf{\em r})$, and for infinitesimal $dt$, with $\textbf{\em r}$ constructed from $\textbf{\em x}$ and $\textbf{\em x}'$ using a parameter $a$, and for the case of a diagonal diffusion matrix, we may write
\begin{align}
  f(\textbf{\em r})&= f((1-a)\textbf{\em x}+a\textbf{\em x}')\nonumber\\
  &\simeq(1-a)f(\textbf{\em x})+af(\textbf{\em x}')\nonumber\\
  &=f(\textbf{\em x})+a(f(\textbf{\em x}')-f(\textbf{\em x}))\nonumber\\
  &=f(\textbf{\em x})+adf(\textbf{\em x})\nonumber\\
  &=f(\textbf{\em x})\nonumber\\
&+a\left(\frac{\partial f(\textbf{\em x})}{\partial t}dt\!+\!\nabla f(\textbf{\em x})\cdot d\textbf{\em x}+\sum_{i}\frac{B_i(\textbf{\em x})^2}{2}\nabla^2f(\textbf{\em x})dt\right)\nonumber\\
  &=f(\textbf{\em x})+a\left(\frac{\partial f(\textbf{\em x})}{\partial t}dt+\sum_{i}\frac{1}{2}B_i(\textbf{\em x})^2\nabla^2f(\textbf{\em x})dt\right.\nonumber\\
&\left.\qquad\qquad+\sum_{i}\frac{\partial f(\textbf{\em x})}{\partial x_i}\left(A_i(\textbf{\em x})dt+B_i(\textbf{\em x})dW_i\right)\right).
\end{align}
Considering all instances of multiplication along with the definitions of $\textbf{\em r}$, $\textbf{\em r}'$, $a$ and $b$, we find the following heuristic rules
\begin{equation}
  f(\textbf{\em r})dt=f(\textbf{\em x})dt+O(dt^{3/2})
\label{heur1}
\end{equation}
\begin{equation}
  f(\textbf{\em r}')dt=f(\textbf{\em x})dt+O(dt^{3/2})
\label{heur2}
\end{equation}
\begin{equation}
  f(\textbf{\em r})dx_i=f(\textbf{\em x})dx_i+2aD_{i}(\textbf{\em x})\frac{\partial f(\textbf{\em x})}{\partial x_i}dt+O(dt^{3/2})
\label{heur3}
\end{equation}
\begin{equation}
  f(\textbf{\em r}')dx_i=f(\textbf{\em x})dx_i+2(1-b)D_{i}(\textbf{\em x})\frac{\partial f(\textbf{\em x})}{\partial x_i}dt+O(dt^{3/2})
\label{heur4}
\end{equation}
giving us a method for converting all multiplications into Ito form. We use a similar reasoning to approximate
\begin{align}
  D_i(\textbf{\em r})&\simeq D_i(\textbf{\em x})+ad(D_i(\textbf{\em x}))\\
  D_i(\textbf{\em r}')&\simeq D_i(\textbf{\em x})+(1-b)d(D_i(\textbf{\em x}))
\end{align}
which along with the approximations to \emph{second} order in $dx_i$, and therefore $d(D_i(\textbf{\em x}))$, of the following form:
\begin{align}
  (1+d(D_i(\textbf{\em x})))^{-1}&\simeq1-d(D_i(\textbf{\em x}))+d(D_i(\textbf{\em x}))^2\\
  \ln(1+d(D_i(\textbf{\em x})))&\simeq d(D_i(\textbf{\em x}))-\frac{d(D_i(\textbf{\em x}))^2}{2}
\end{align}
and an Ito definition of $d(D(\textbf{\em x}))$, allow us to write the first four terms in Eq.~(\ref{ratiofull}) to first order in $dt$ as
\begin{align}
 \frac{1}{2}&\ln{D_{i}(\textbf{\em r}')}\!-\!\frac{1}{2}\ln{D_{i}(\textbf{\em r})}\!+\!\frac{dx_i^2}{4D_{i}(\textbf{\em r}')dt}\!-\!\frac{dx_i^2}{4D_{i}(\textbf{\em r})dt}\nonumber\\
&\simeq\frac{((1-b)^2-a^2)}{2D_{i}(\textbf{\em x})}\left(\frac{\partial D_{i}(\textbf{\em x})}{\partial x_i}\right)^2dt.
\end{align}
Using the above, and the heuristic rules in Eqs.~(\ref{heur1}-\ref{heur4}), we obtain
\begin{align}
  d\Delta S_{\rm med}&=\sum_{i}\frac{A_i^{\rm ir}(\textbf{\em x})}{D_{i}(\textbf{\em x})}dx_i-\frac{A^{\rm rev}_i(\textbf{\em x})A^{\rm ir}_i(\textbf{\em x})}{D_{i}(\textbf{\em x})}dt\nonumber\\
&+\frac{\partial A^{\rm ir}_i(\textbf{\em x})}{\partial x_i}dt-\frac{\partial A^{\rm rev}_i(\textbf{\em x})}{\partial x_i}dt\nonumber\\
&+\!\frac{A^{\rm rev}_i(\textbf{\em x})}{D_{i}(\textbf{\em x})}\frac{\partial D_{i}(\textbf{\em x})}{\partial x_i}dt\!-\!\frac{A^{\rm ir}_i(\textbf{\em x})}{D_{i}(\textbf{\em x})}\frac{\partial D_{i}(\textbf{\em x})}{\partial x_i}dt\nonumber\\
&-(a\!+\!b)\frac{1}{D_{i}(\textbf{\em x})}\frac{\partial D_{i}(\textbf{\em x})}{\partial x_i}dx_i\nonumber\\
&\!\!\!\!\!\!\!\!\!\!\!\!+(b^2\!-2b\!-\!a^2)\left(\frac{\partial^2 D_{i}(\textbf{\em x})}{\partial x_i^2}-\frac{1}{D_{i}(\textbf{\em x})}\left(\frac{\partial D_{i}(\textbf{\em x})}{\partial x_i}\right)^2\right)dt\nonumber\\
&+\frac{((1\!-\!b)^2\!-\!a^2)}{2D_{i}(\textbf{\em x})}\left(\frac{\partial D_{i}(\textbf{\em x})}{\partial x_i}\right)^2dt
\end{align}
which depends on the choice of $a$ and $b$. We note, however that without multiplicative noise (ie. for $(\partial/\partial x_i)D_{i}(\textbf{\em x})=0$), where the inherent mathematical ambiguity in stochastic integrals is absent, the dependence on evaluation points ($a$ and $b$) disappears. So this dependence is evidently related to the ambiguity of evaluation point in a stochastic integral. However, the underlying SDEs and entropy are not, and should not be, ambiguous since we have specified Ito SDEs and have used the short time propagator appropriate for their corresponding Fokker-Planck equation. Since all evaluation points lead to the correct path probability density we are not obliged to consider, for example, only Ito-type multiplication ($a=b=0$) simply because the underlying SDEs are of Ito form. Rather, to proceed we recognise that with multiplicative noise we must ensure that we evaluate the two transition probability densities at precisely the same coordinates  and not just at same time (which would suffice for additive noise), for the same reasons that make stochastic integration sensitive to the specific integration scheme when the integrand has dependence on the integrating variable. Alternatively it may be reasoned that as $dt\to 0$ we require the short time propagators to approach jump transition probabilities of a master equation. Under such a description the entropy production can be unambiguously described by the ratios of probabilitities appearing in a master equation approach \cite{adiabaticnonadiabatic0,prl}. Therefore, if we represent such a quantity using the short time propagators we require the transition rates in both numerator and denominator to be evaluated equivalently. Since these are characterised by our system variables $A_i$, $B_i$ etc, to effect such a condition we require $\textbf{\em r}'=\textbf{\em r}$ which is equivalent to making the choice $b=1-a$, noting that $a$ is still a free parameter. One may think of this as insisting that the path transformation $\vec{\textbf{\em x}}^{\dagger}(t)= \boldsymbol{\varepsilon}\vec{\textbf{\em x}}(\tau-t)$ persists on a (sub) infinitesimal scale, which would not matter in normal calculus, so that the noise is experienced in precisely the right way. Inserting $b=1-a$ into the above yields Eq.~(\ref{itomed}) which has no dependence on the choice $a$, an indication that it is defined in a sound fashion.
\section{Consistency of entropy contributions}
\label{appendixB}
By construction, we have $\Delta S_{\rm tot}=\Delta S_1+\Delta S_2+\Delta S_3$ which according to the expressions in Eqs.~(\ref{StotAV}), (\ref{S1av}), (\ref{S2av}) and (\ref{S3av}) means we require
\begin{align}
\frac{d\langle \Delta S_{\rm tot}\rangle}{dt}&=\sum_{i}\int d\textbf{\em x}\;\frac{(J^{\rm ir}_i(\textbf{\em x}))^2}{p(\textbf{\em x})D_i(\textbf{\em x})}\nonumber\\
&=\sum_i\int d\textbf{\em x}\;\frac{ p(\textbf{\em x})}{D_i(\textbf{\em x})}\left(\frac{J^{\rm ir}_i(\textbf{\em x})}{p(\textbf{\em x})}-\frac{J^{\rm st,ir}_i(\textbf{\em x})}{p^{\rm st}(\textbf{\em x})}\right)^2\nonumber\\
&\quad+\int d\textbf{\em x}\; \frac{p(\textbf{\em x})}{D_i(\textbf{\em x})}\left(\frac{J^{\rm ir,st}_i(\boldsymbol{\varepsilon}\textbf{\em x})}{p^{\rm st}(\boldsymbol{\varepsilon}\textbf{\em x})}\right)^2\nonumber\\
&\quad+\int d\textbf{\em x}\; J_i(\textbf{\em x})(\phi_i'(\textbf{\em x})-\varepsilon_i\phi_i'(\boldsymbol{\varepsilon}\textbf{\em x})).
\end{align}
Thus in order that everything is consistent we require
\begin{align}
  0&=\sum_i\int d\textbf{\em x}\Bigg[-2\frac{J^{\rm ir}_i(\textbf{\em x})}{D_i(\textbf{\em x})}\frac{J^{\rm ir,st}_i(\textbf{\em x})}{p^{\rm st}(\textbf{\em x})}+\frac{p(\textbf{\em x})}{D_i(\textbf{\em x})}\left(\frac{J^{\rm ir,st}_i(\textbf{\em x})}{p^{\rm st}(\textbf{\em x})}\right)^2\nonumber\\
&+\frac{p(\textbf{\em x})}{D_i(\textbf{\em x})}\left(\frac{J^{\rm ir,st}_i(\boldsymbol{\varepsilon}\textbf{\em x})}{p^{\rm st}(\boldsymbol{\varepsilon}\textbf{\em x})}\right)^2+J_i(\textbf{\em x})(\phi_i'(\textbf{\em x})-\varepsilon_i\phi_i'(\boldsymbol{\varepsilon}\textbf{\em x}))\Bigg].
\end{align}
By substitution of the definitions of the fluxes this reduces to
\begin{align}
  0=\sum_i\int d\textbf{\em x}\;\Bigg[&2A^{\rm ir}_i(\textbf{\em x})\frac{\partial p(\textbf{\em x})}{\partial x_i}-2\frac{\partial D_i(\textbf{\em x})}{\partial x_i}\frac{\partial p(\textbf{\em x})}{\partial x_i}\nonumber\\
&+\frac{\partial(D_i(\textbf{\em x})p(\textbf{\em x}))}{\partial x_i}\left(\phi_i'(\textbf{\em x})+\varepsilon_i\phi_i'(\boldsymbol{\varepsilon}\textbf{\em x})\right)\nonumber\\
&+p(\textbf{\em x})D_i(\textbf{\em x})\left((\phi_i'(\textbf{\em x}))^2+(\phi_i'(\boldsymbol{\varepsilon}\textbf{\em x}))^2\right)\nonumber\\
&-2p(\textbf{\em x})\frac{\partial D_i(\textbf{\em x})}{\partial x_i}\left(\phi_i'(\textbf{\em x})+\varepsilon_i\phi_i'(\boldsymbol{\varepsilon}\textbf{\em x})\right)\nonumber\\
&+p(\textbf{\em x})(A^{\rm ir}_i(\textbf{\em x})+A^{\rm rev}_i(\textbf{\em x}))\phi_i'(\textbf{\em x})\nonumber\\
&+\varepsilon_ip(\textbf{\em x})(A^{\rm ir}_i(\textbf{\em x})-A^{\rm rev}_i(\textbf{\em x}))\phi_i'(\boldsymbol{\varepsilon}\textbf{\em x})\Bigg].
\end{align}
Integrating by parts, dropping or cancelling boundary terms and using the definition of the irreversible and reversible drift terms yields the condition
\begin{align}
  &0=\sum_i\int d\textbf{\em x}\;p(\textbf{\em x})\Bigg[-2\frac{\partial A^{\rm ir}_i(\textbf{\em x})}{\partial x_i}+2\frac{\partial^2 D_i(\textbf{\em x})}{\partial x^2_i}\nonumber\\
&-D_i(\textbf{\em x})\left(\phi_i''(\textbf{\em x})+\phi_i''(\boldsymbol{\varepsilon}\textbf{\em x})\right)+D_i(\textbf{\em x})\left((\phi_i'(\textbf{\em x}))^2+(\phi_i'(\boldsymbol{\varepsilon}\textbf{\em x}))^2\right)\nonumber\\
&-2\frac{\partial D_i(\textbf{\em x})}{\partial x_i}\left(\phi_i'(\textbf{\em x})+\varepsilon_i\phi_i'(\boldsymbol{\varepsilon}\textbf{\em x})\right)\nonumber+(A^{\rm ir}_i(\textbf{\em x})+A^{\rm rev}_i(\textbf{\em x}))\phi_i'(\textbf{\em x})\nonumber\\
&\qquad\qquad+(A^{\rm ir}_i(\boldsymbol{\varepsilon}\textbf{\em x})+A^{\rm rev}_i(\boldsymbol{\varepsilon}\textbf{\em x}))\phi_i'(\boldsymbol{\varepsilon}\textbf{\em x})\Bigg].
 \label{sum0}
\end{align}
The divergenceless stationary distribution condition however yields
\begin{align}
 0=\Bigg(&-(A^{\rm ir}_i(\textbf{\em x})+A^{\rm rev}_i(\textbf{\em x}))\phi_i'(\textbf{\em x})-D_i(\textbf{\em x})(\phi_i'(\textbf{\em x}))^2\nonumber\\
&+\frac{\partial A^{\rm ir}_i(\textbf{\em x})}{\partial x_i}+\frac{\partial A^{\rm rev}_i(\textbf{\em x})}{\partial x_i}-\frac{\partial^2D_i(\textbf{\em x})}{\partial x_i^2}\nonumber\\
&\quad+D_i(\textbf{\em x})\phi_i''(\textbf{\em x})+2\frac{\partial D_i(\textbf{\em x})}{\partial x_i}\phi_i'(\textbf{\em x})\Bigg)e^{-\phi(\textbf{\em x})},
\label{div1}
\end{align}
but also
\begin{align}
 0=\Bigg(&-(A^{\rm ir}_i(\boldsymbol{\varepsilon}\textbf{\em x})+A^{\rm rev}_i(\boldsymbol{\varepsilon}\textbf{\em x}))\phi_i'(\boldsymbol{\varepsilon}\textbf{\em x})-D_i(\boldsymbol{\varepsilon}\textbf{\em x})(\phi_i'(\boldsymbol{\varepsilon}\textbf{\em x}))^2\nonumber\\
&+\frac{\partial A^{\rm ir}_i(\boldsymbol{\varepsilon}\textbf{\em x})}{\partial (\varepsilon_ix_i)}+\frac{\partial A^{\rm rev}_i(\boldsymbol{\varepsilon}\textbf{\em x})}{\partial (\varepsilon_ix_i)}-\frac{\partial^2D_i(\boldsymbol{\varepsilon}\textbf{\em x})}{\partial (\varepsilon_ix_i)^2}\nonumber\\
&+D_i(\boldsymbol{\varepsilon}\textbf{\em x})\phi_i''(\boldsymbol{\varepsilon}\textbf{\em x})+2\frac{\partial D_i(\boldsymbol{\varepsilon}\textbf{\em x})}{\partial (\varepsilon_ix_i)}\phi_i'(\boldsymbol{\varepsilon}\textbf{\em x})\Bigg)e^{-\phi(\boldsymbol{\varepsilon}\textbf{\em x})}.
\label{div2}
\end{align}
Combining the two conditions in Eqs.~(\ref{div1}) and (\ref{div2}) yields the contents of the brackets in equation (\ref{sum0}) and so the result is proved.
\providecommand{\noopsort}[1]{}\providecommand{\singleletter}[1]{#1}%


\begin{thebibliography}{38}%
\makeatletter
\providecommand \@ifxundefined [1]{%
 \@ifx{#1\undefined}
}%
\providecommand \@ifnum [1]{%
 \ifnum #1\expandafter \@firstoftwo
 \else \expandafter \@secondoftwo
 \fi
}%
\providecommand \@ifx [1]{%
 \ifx #1\expandafter \@firstoftwo
 \else \expandafter \@secondoftwo
 \fi
}%
\providecommand \natexlab [1]{#1}%
\providecommand \enquote  [1]{``#1''}%
\providecommand \bibnamefont  [1]{#1}%
\providecommand \bibfnamefont [1]{#1}%
\providecommand \citenamefont [1]{#1}%
\providecommand \href@noop [0]{\@secondoftwo}%
\providecommand \href [0]{\begingroup \@sanitize@url \@href}%
\providecommand \@href[1]{\@@startlink{#1}\@@href}%
\providecommand \@@href[1]{\endgroup#1\@@endlink}%
\providecommand \@sanitize@url [0]{\catcode `\\12\catcode `\$12\catcode
  `\&12\catcode `\#12\catcode `\^12\catcode `\_12\catcode `\%12\relax}%
\providecommand \@@startlink[1]{}%
\providecommand \@@endlink[0]{}%
\providecommand \url  [0]{\begingroup\@sanitize@url \@url }%
\providecommand \@url [1]{\endgroup\@href {#1}{\urlprefix }}%
\providecommand \urlprefix  [0]{URL }%
\providecommand \Eprint [0]{\href }%
\providecommand \doibase [0]{http://dx.doi.org/}%
\providecommand \selectlanguage [0]{\@gobble}%
\providecommand \bibinfo  [0]{\@secondoftwo}%
\providecommand \bibfield  [0]{\@secondoftwo}%
\providecommand \translation [1]{[#1]}%
\providecommand \BibitemOpen [0]{}%
\providecommand \bibitemStop [0]{}%
\providecommand \bibitemNoStop [0]{.\EOS\space}%
\providecommand \EOS [0]{\spacefactor3000\relax}%
\providecommand \BibitemShut  [1]{\csname bibitem#1\endcsname}%
\let\auto@bib@innerbib\@empty
\bibitem [{\citenamefont {Evans}\ \emph {et~al.}(1993)\citenamefont {Evans},
  \citenamefont {Cohen},\ and\ \citenamefont {Morriss}}]{Evans93}%
  \BibitemOpen
  \bibfield  {author} {\bibinfo {author} {\bibfnamefont {D.~J.}\ \bibnamefont
  {Evans}}, \bibinfo {author} {\bibfnamefont {E.~G.~D.}\ \bibnamefont {Cohen}},
  \ and\ \bibinfo {author} {\bibfnamefont {G.~P.}\ \bibnamefont {Morriss}},\
  }\href@noop {} {\bibfield  {journal} {\bibinfo  {journal} {Phys. Rev. Lett.}\
  }\textbf {\bibinfo {volume} {71}},\ \bibinfo {pages} {2401} (\bibinfo {year}
  {1993})}\BibitemShut {NoStop}%
\bibitem [{\citenamefont {Evans}\ and\ \citenamefont
  {Searles}(1995)}]{Evans95}%
  \BibitemOpen
  \bibfield  {author} {\bibinfo {author} {\bibfnamefont {D.~J.}\ \bibnamefont
  {Evans}}\ and\ \bibinfo {author} {\bibfnamefont {D.~J.}\ \bibnamefont
  {Searles}},\ }\href@noop {} {\bibfield  {journal} {\bibinfo  {journal} {Phys.
  Rev. E}\ }\textbf {\bibinfo {volume} {52}},\ \bibinfo {pages} {5839}
  (\bibinfo {year} {1995})}\BibitemShut {NoStop}%
\bibitem [{\citenamefont {Evans}\ and\ \citenamefont
  {Searles}(2002)}]{Evans02}%
  \BibitemOpen
  \bibfield  {author} {\bibinfo {author} {\bibfnamefont {D.~J.}\ \bibnamefont
  {Evans}}\ and\ \bibinfo {author} {\bibfnamefont {D.~J.}\ \bibnamefont
  {Searles}},\ }\href@noop {} {\bibfield  {journal} {\bibinfo  {journal} {Adv.
  Phys.}\ }\textbf {\bibinfo {volume} {51}},\ \bibinfo {pages} {1529} (\bibinfo
  {year} {2002})}\BibitemShut {NoStop}%
\bibitem [{\citenamefont {Carberry}\ \emph {et~al.}(2004)\citenamefont
  {Carberry}, \citenamefont {Reid}, \citenamefont {G.M.Wang}, \citenamefont
  {Sevick}, \citenamefont {Searles},\ and\ \citenamefont {Evans}}]{Carberry04}%
  \BibitemOpen
  \bibfield  {author} {\bibinfo {author} {\bibfnamefont {D.}~\bibnamefont
  {Carberry}}, \bibinfo {author} {\bibfnamefont {J.~C.}\ \bibnamefont {Reid}},
  \bibinfo {author} {\bibnamefont {G.M.Wang}}, \bibinfo {author} {\bibfnamefont
  {E.}~\bibnamefont {Sevick}}, \bibinfo {author} {\bibfnamefont {D.~J.}\
  \bibnamefont {Searles}}, \ and\ \bibinfo {author} {\bibfnamefont {D.~J.}\
  \bibnamefont {Evans}},\ }\href@noop {} {\bibfield  {journal} {\bibinfo
  {journal} {Phys. Rev. Lett.}\ }\textbf {\bibinfo {volume} {92}},\ \bibinfo
  {pages} {140601} (\bibinfo {year} {2004})}\BibitemShut {NoStop}%
\bibitem [{\citenamefont {Gallavotti}\ and\ \citenamefont
  {Cohen}(1995)}]{Gallavotti95}%
  \BibitemOpen
  \bibfield  {author} {\bibinfo {author} {\bibfnamefont {G.}~\bibnamefont
  {Gallavotti}}\ and\ \bibinfo {author} {\bibfnamefont {E.~G.~D.}\ \bibnamefont
  {Cohen}},\ }\href@noop {} {\bibfield  {journal} {\bibinfo  {journal} {Phys.
  Rev. Lett.}\ }\textbf {\bibinfo {volume} {74}},\ \bibinfo {pages} {2694}
  (\bibinfo {year} {1995})}\BibitemShut {NoStop}%
\bibitem [{\citenamefont {Kurchan}(1998)}]{kurchan}%
  \BibitemOpen
  \bibfield  {author} {\bibinfo {author} {\bibfnamefont {J.}~\bibnamefont
  {Kurchan}},\ }\href@noop {} {\bibfield  {journal} {\bibinfo  {journal} {J.
  Phys. A: Math. Gen.}\ }\textbf {\bibinfo {volume} {31}},\ \bibinfo {pages}
  {3719} (\bibinfo {year} {1998})}\BibitemShut {NoStop}%
\bibitem [{\citenamefont {Lebowitz}\ and\ \citenamefont
  {Spohn}(1999)}]{GCforstochastic}%
  \BibitemOpen
  \bibfield  {author} {\bibinfo {author} {\bibfnamefont {J.~L.}\ \bibnamefont
  {Lebowitz}}\ and\ \bibinfo {author} {\bibfnamefont {H.}~\bibnamefont
  {Spohn}},\ }\href@noop {} {\bibfield  {journal} {\bibinfo  {journal} {J.
  Stat. Phys.}\ }\textbf {\bibinfo {volume} {95}},\ \bibinfo {pages} {333}
  (\bibinfo {year} {1999})}\BibitemShut {NoStop}%
\bibitem [{\citenamefont {Jarzynski}(1997)}]{Jarzynski97}%
  \BibitemOpen
  \bibfield  {author} {\bibinfo {author} {\bibfnamefont {C.}~\bibnamefont
  {Jarzynski}},\ }\href@noop {} {\bibfield  {journal} {\bibinfo  {journal}
  {Phys. Rev. Lett.}\ }\textbf {\bibinfo {volume} {78}},\ \bibinfo {pages}
  {2690} (\bibinfo {year} {1997})}\BibitemShut {NoStop}%
\bibitem [{\citenamefont {Crooks}(1999)}]{crooksoriginal}%
  \BibitemOpen
  \bibfield  {author} {\bibinfo {author} {\bibfnamefont {G.~E.}\ \bibnamefont
  {Crooks}},\ }\href@noop {} {\bibfield  {journal} {\bibinfo  {journal} {Phys.
  Rev. E}\ }\textbf {\bibinfo {volume} {60}},\ \bibinfo {pages} {2721}
  (\bibinfo {year} {1999})}\BibitemShut {NoStop}%
\bibitem [{\citenamefont {Crooks}(1998)}]{Crooks98}%
  \BibitemOpen
  \bibfield  {author} {\bibinfo {author} {\bibfnamefont {G.~E.}\ \bibnamefont
  {Crooks}},\ }\href@noop {} {\bibfield  {journal} {\bibinfo  {journal} {J.
  Stat. Phys.}\ }\textbf {\bibinfo {volume} {90}},\ \bibinfo {pages} {1481}
  (\bibinfo {year} {1998})}\BibitemShut {NoStop}%
\bibitem [{\citenamefont {Seifert}(2005)}]{seifertoriginal}%
  \BibitemOpen
  \bibfield  {author} {\bibinfo {author} {\bibfnamefont {U.}~\bibnamefont
  {Seifert}},\ }\href@noop {} {\bibfield  {journal} {\bibinfo  {journal} {Phys.
  Rev. Lett.}\ }\textbf {\bibinfo {volume} {95}},\ \bibinfo {pages} {040602}
  (\bibinfo {year} {2005})}\BibitemShut {NoStop}%
\bibitem [{\citenamefont {Sekimoto}(1998{\natexlab{a}})}]{sekimoto}%
  \BibitemOpen
  \bibfield  {author} {\bibinfo {author} {\bibfnamefont {K.}~\bibnamefont
  {Sekimoto}},\ }\href@noop {} {\bibfield  {journal} {\bibinfo  {journal}
  {Prog. Theor. Phys. Suppl.}\ }\textbf {\bibinfo {volume} {130}},\ \bibinfo
  {pages} {17} (\bibinfo {year} {1998}{\natexlab{a}})}\BibitemShut {NoStop}%
\bibitem [{\citenamefont {Hatano}\ and\ \citenamefont
  {Sasa}(2001)}]{hatanosasa}%
  \BibitemOpen
  \bibfield  {author} {\bibinfo {author} {\bibfnamefont {T.}~\bibnamefont
  {Hatano}}\ and\ \bibinfo {author} {\bibfnamefont {S.}~\bibnamefont {Sasa}},\
  }\href@noop {} {\bibfield  {journal} {\bibinfo  {journal} {Phys. Rev. Lett.}\
  }\textbf {\bibinfo {volume} {86}},\ \bibinfo {pages} {3463} (\bibinfo {year}
  {2001})}\BibitemShut {NoStop}%
\bibitem [{\citenamefont {Speck}\ and\ \citenamefont
  {Seifert}(2005)}]{IFThousekeeping}%
  \BibitemOpen
  \bibfield  {author} {\bibinfo {author} {\bibfnamefont {T.}~\bibnamefont
  {Speck}}\ and\ \bibinfo {author} {\bibfnamefont {U.}~\bibnamefont
  {Seifert}},\ }\href@noop {} {\bibfield  {journal} {\bibinfo  {journal} {J.
  Phys. A: Math. Gen.}\ }\textbf {\bibinfo {volume} {38}},\ \bibinfo {pages}
  {L581} (\bibinfo {year} {2005})}\BibitemShut {NoStop}%
\bibitem [{\citenamefont {Chernyak}\ \emph {et~al.}(2006)\citenamefont
  {Chernyak}, \citenamefont {Chertkov},\ and\ \citenamefont
  {Jarzynski}}]{Jarpathintegral}%
  \BibitemOpen
  \bibfield  {author} {\bibinfo {author} {\bibfnamefont {V.~Y.}\ \bibnamefont
  {Chernyak}}, \bibinfo {author} {\bibfnamefont {M.}~\bibnamefont {Chertkov}},
  \ and\ \bibinfo {author} {\bibfnamefont {C.}~\bibnamefont {Jarzynski}},\
  }\href@noop {} {\bibfield  {journal} {\bibinfo  {journal} {J. Stat. Mech.
  \textup{P08001}}\ } (\bibinfo {year} {2006})}\BibitemShut {NoStop}%
\bibitem [{\citenamefont {Esposito}\ \emph {et~al.}(2007)\citenamefont
  {Esposito}, \citenamefont {Harbola},\ and\ \citenamefont
  {Mukamel}}]{Esposito07}%
  \BibitemOpen
  \bibfield  {author} {\bibinfo {author} {\bibfnamefont {M.}~\bibnamefont
  {Esposito}}, \bibinfo {author} {\bibfnamefont {U.}~\bibnamefont {Harbola}}, \
  and\ \bibinfo {author} {\bibfnamefont {S.}~\bibnamefont {Mukamel}},\ }\href
  {\doibase 10.1103/PhysRevE.76.031132} {\bibfield  {journal} {\bibinfo
  {journal} {Phys. Rev. E}\ }\textbf {\bibinfo {volume} {76}},\ \bibinfo
  {pages} {031132} (\bibinfo {year} {2007})}\BibitemShut {NoStop}%
\bibitem [{\citenamefont {Ge}(2009)}]{Ge09}%
  \BibitemOpen
  \bibfield  {author} {\bibinfo {author} {\bibfnamefont {H.}~\bibnamefont
  {Ge}},\ }\href {\doibase 10.1103/PhysRevE.80.021137} {\bibfield  {journal}
  {\bibinfo  {journal} {Phys. Rev. E}\ }\textbf {\bibinfo {volume} {80}},\
  \bibinfo {pages} {021137} (\bibinfo {year} {2009})}\BibitemShut {NoStop}%
\bibitem [{\citenamefont {Ge}\ and\ \citenamefont {Qian}(2010)}]{Ge10}%
  \BibitemOpen
  \bibfield  {author} {\bibinfo {author} {\bibfnamefont {H.}~\bibnamefont
  {Ge}}\ and\ \bibinfo {author} {\bibfnamefont {H.}~\bibnamefont {Qian}},\
  }\href {\doibase 10.1103/PhysRevE.81.051133} {\bibfield  {journal} {\bibinfo
  {journal} {Phys. Rev. E}\ }\textbf {\bibinfo {volume} {81}},\ \bibinfo
  {pages} {051133} (\bibinfo {year} {2010})}\BibitemShut {NoStop}%
\bibitem [{\citenamefont {Esposito}\ and\ \citenamefont {Van~den
  Broeck}(2010{\natexlab{a}})}]{adiabaticnonadiabatic0}%
  \BibitemOpen
  \bibfield  {author} {\bibinfo {author} {\bibfnamefont {M.}~\bibnamefont
  {Esposito}}\ and\ \bibinfo {author} {\bibfnamefont {C.}~\bibnamefont {Van~den
  Broeck}},\ }\href@noop {} {\bibfield  {journal} {\bibinfo  {journal} {Phys.
  Rev. Lett.}\ }\textbf {\bibinfo {volume} {104}},\ \bibinfo {pages} {090601}
  (\bibinfo {year} {2010}{\natexlab{a}})}\BibitemShut {NoStop}%
\bibitem [{\citenamefont {Esposito}\ and\ \citenamefont {Van~den
  Broeck}(2010{\natexlab{b}})}]{adiabaticnonadiabatic1}%
  \BibitemOpen
  \bibfield  {author} {\bibinfo {author} {\bibfnamefont {M.}~\bibnamefont
  {Esposito}}\ and\ \bibinfo {author} {\bibfnamefont {C.}~\bibnamefont {Van~den
  Broeck}},\ }\href@noop {} {\bibfield  {journal} {\bibinfo  {journal} {Phys.
  Rev. E}\ }\textbf {\bibinfo {volume} {82}},\ \bibinfo {pages} {011143}
  (\bibinfo {year} {2010}{\natexlab{b}})}\BibitemShut {NoStop}%
\bibitem [{\citenamefont {Van~den Broeck}\ and\ \citenamefont
  {Esposito}(2010)}]{adiabaticnonadiabatic2}%
  \BibitemOpen
  \bibfield  {author} {\bibinfo {author} {\bibfnamefont {C.}~\bibnamefont
  {Van~den Broeck}}\ and\ \bibinfo {author} {\bibfnamefont {M.}~\bibnamefont
  {Esposito}},\ }\href@noop {} {\bibfield  {journal} {\bibinfo  {journal}
  {Phys. Rev. E}\ }\textbf {\bibinfo {volume} {82}},\ \bibinfo {pages} {011144}
  (\bibinfo {year} {2010})}\BibitemShut {NoStop}%
\bibitem [{\citenamefont {Oono}\ and\ \citenamefont {Paniconi}(1998)}]{oono}%
  \BibitemOpen
  \bibfield  {author} {\bibinfo {author} {\bibfnamefont {Y.}~\bibnamefont
  {Oono}}\ and\ \bibinfo {author} {\bibfnamefont {M.}~\bibnamefont
  {Paniconi}},\ }\href@noop {} {\bibfield  {journal} {\bibinfo  {journal}
  {Prog. Theor. Phys. Suppl.}\ }\textbf {\bibinfo {volume} {130}},\ \bibinfo
  {pages} {29} (\bibinfo {year} {1998})}\BibitemShut {NoStop}%
\bibitem [{\citenamefont {Spinney}\ and\ \citenamefont {Ford}()}]{prl}%
  \BibitemOpen
  \bibfield  {author} {\bibinfo {author} {\bibfnamefont {R.~E.}\ \bibnamefont
  {Spinney}}\ and\ \bibinfo {author} {\bibfnamefont {I.~J.}\ \bibnamefont
  {Ford}},\ }\href@noop {} {\bibinfo  {journal} {arXiv:1201.0904}\
  }\BibitemShut {NoStop}%
\bibitem [{\citenamefont {Harris}\ and\ \citenamefont
  {Sch{\"u}tz}(2007)}]{Harris07}%
  \BibitemOpen
\bibfield  {journal} {  }\bibfield  {author} {\bibinfo {author} {\bibfnamefont
  {R.~J.}\ \bibnamefont {Harris}}\ and\ \bibinfo {author} {\bibfnamefont
  {G.~M.}\ \bibnamefont {Sch{\"u}tz}},\ }\href@noop {} {\bibfield  {journal}
  {\bibinfo  {journal} {J. Stat. Mech. \textup{P07020}}\ } (\bibinfo {year}
  {2007})}\BibitemShut {NoStop}%
\bibitem [{\citenamefont {Seifert}(2008)}]{seifertprinciples}%
  \BibitemOpen
  \bibfield  {author} {\bibinfo {author} {\bibfnamefont {U.}~\bibnamefont
  {Seifert}},\ }\href@noop {} {\bibfield  {journal} {\bibinfo  {journal} {Eur.
  Phys. J. B}\ }\textbf {\bibinfo {volume} {64}},\ \bibinfo {pages} {423}
  (\bibinfo {year} {2008})}\BibitemShut {NoStop}%
\bibitem [{\citenamefont {Wissel}(1979)}]{shorttimepropagator}%
  \BibitemOpen
  \bibfield  {author} {\bibinfo {author} {\bibfnamefont {C.}~\bibnamefont
  {Wissel}},\ }\href@noop {} {\bibfield  {journal} {\bibinfo  {journal} {Z.
  Physik B}\ }\textbf {\bibinfo {volume} {35}},\ \bibinfo {pages} {185}
  (\bibinfo {year} {1979})}\BibitemShut {NoStop}%
\bibitem [{\citenamefont {Tom\'e}\ and\ \citenamefont
  {de~Oliveira}(2010)}]{Tome}%
  \BibitemOpen
  \bibfield  {author} {\bibinfo {author} {\bibfnamefont {T.}~\bibnamefont
  {Tom\'e}}\ and\ \bibinfo {author} {\bibfnamefont {M.~J.}\ \bibnamefont
  {de~Oliveira}},\ }\href@noop {} {\bibfield  {journal} {\bibinfo  {journal}
  {Phys. Rev. E}\ }\textbf {\bibinfo {volume} {82}},\ \bibinfo {pages} {021120}
  (\bibinfo {year} {2010})}\BibitemShut {NoStop}%
\bibitem [{\citenamefont {Daems}\ and\ \citenamefont
  {Nicolis}(1999)}]{positivewrong}%
  \BibitemOpen
  \bibfield  {author} {\bibinfo {author} {\bibfnamefont {D.}~\bibnamefont
  {Daems}}\ and\ \bibinfo {author} {\bibfnamefont {G.}~\bibnamefont
  {Nicolis}},\ }\href@noop {} {\bibfield  {journal} {\bibinfo  {journal} {Phys.
  Rev. E}\ }\textbf {\bibinfo {volume} {59}},\ \bibinfo {pages} {4000}
  (\bibinfo {year} {1999})}\BibitemShut {NoStop}%
\bibitem [{\citenamefont {Matsuo}\ and\ \citenamefont
  {Sasa}(2000)}]{itostrat1}%
  \BibitemOpen
  \bibfield  {author} {\bibinfo {author} {\bibfnamefont {M.}~\bibnamefont
  {Matsuo}}\ and\ \bibinfo {author} {\bibfnamefont {S.}~\bibnamefont {Sasa}},\
  }\href@noop {} {\bibfield  {journal} {\bibinfo  {journal} {Physica A}\
  }\textbf {\bibinfo {volume} {276}},\ \bibinfo {pages} {188 } (\bibinfo {year}
  {2000})}\BibitemShut {NoStop}%
\bibitem [{\citenamefont {Kampen}(1988)}]{itostrat2}%
  \BibitemOpen
  \bibfield  {author} {\bibinfo {author} {\bibfnamefont {N.~V.}\ \bibnamefont
  {Kampen}},\ }\href@noop {} {\bibfield  {journal} {\bibinfo  {journal} {J.
  Phys. Chem. Sol.}\ }\textbf {\bibinfo {volume} {49}},\ \bibinfo {pages} {673
  } (\bibinfo {year} {1988})}\BibitemShut {NoStop}%
\bibitem [{\citenamefont {Stolovitzky}(1998)}]{stol}%
  \BibitemOpen
  \bibfield  {author} {\bibinfo {author} {\bibfnamefont {G.}~\bibnamefont
  {Stolovitzky}},\ }\href@noop {} {\bibfield  {journal} {\bibinfo  {journal}
  {Phys. Lett. A}\ }\textbf {\bibinfo {volume} {241}},\ \bibinfo {pages} {240 }
  (\bibinfo {year} {1998})}\BibitemShut {NoStop}%
\bibitem [{\citenamefont {Sekimoto}(2010)}]{sekimotobook}%
  \BibitemOpen
  \bibfield  {author} {\bibinfo {author} {\bibfnamefont {K.}~\bibnamefont
  {Sekimoto}},\ }\href@noop {} {\emph {\bibinfo {title} {Stochastic
  Energetics}}},\ \bibinfo {series} {Lecture Notes in Physics}, Vol.\ \bibinfo
  {volume} {799}\ (\bibinfo  {publisher} {Springer},\ \bibinfo {address}
  {Berlin Heidelberg},\ \bibinfo {year} {2010})\BibitemShut {NoStop}%
\bibitem [{\citenamefont {Widder}\ and\ \citenamefont
  {Titulaer}(1989)}]{Widder}%
  \BibitemOpen
  \bibfield  {author} {\bibinfo {author} {\bibfnamefont {M.}~\bibnamefont
  {Widder}}\ and\ \bibinfo {author} {\bibfnamefont {U.}~\bibnamefont
  {Titulaer}},\ }\href@noop {} {\bibfield  {journal} {\bibinfo  {journal}
  {Physica A}\ }\textbf {\bibinfo {volume} {154}},\ \bibinfo {pages} {452 }
  (\bibinfo {year} {1989})}\BibitemShut {NoStop}%
\bibitem [{\citenamefont {Chaikin}\ and\ \citenamefont
  {Lubensky}(1995)}]{chaikin}%
  \BibitemOpen
  \bibfield  {author} {\bibinfo {author} {\bibfnamefont {P.~M.}\ \bibnamefont
  {Chaikin}}\ and\ \bibinfo {author} {\bibfnamefont {T.~C.}\ \bibnamefont
  {Lubensky}},\ }\href@noop {} {\emph {\bibinfo {title} {Principles of
  Condensed Matter Physics}}}\ (\bibinfo  {publisher} {Cambridge University
  Press},\ \bibinfo {address} {Cambridge},\ \bibinfo {year} {1995})\BibitemShut
  {NoStop}%
\bibitem [{\citenamefont {Sekimoto}(1998{\natexlab{b}})}]{sekimoto1}%
  \BibitemOpen
  \bibfield  {author} {\bibinfo {author} {\bibfnamefont {K.}~\bibnamefont
  {Sekimoto}},\ }\href@noop {} {\bibfield  {journal} {\bibinfo  {journal}
  {Prog. Theor. Phys. Suppl.}\ }\textbf {\bibinfo {volume} {130}},\ \bibinfo
  {pages} {17} (\bibinfo {year} {1998}{\natexlab{b}})}\BibitemShut {NoStop}%
\bibitem [{\citenamefont {Saha}\ \emph {et~al.}(2009)\citenamefont {Saha},
  \citenamefont {Lahiri},\ and\ \citenamefont {Jayannavar}}]{Saha09}%
  \BibitemOpen
  \bibfield  {author} {\bibinfo {author} {\bibfnamefont {A.}~\bibnamefont
  {Saha}}, \bibinfo {author} {\bibfnamefont {S.}~\bibnamefont {Lahiri}}, \ and\
  \bibinfo {author} {\bibfnamefont {A.~M.}\ \bibnamefont {Jayannavar}},\
  }\href@noop {} {\bibfield  {journal} {\bibinfo  {journal} {Phys. Rev. E}\
  }\textbf {\bibinfo {volume} {80}},\ \bibinfo {pages} {011117} (\bibinfo
  {year} {2009})}\BibitemShut {NoStop}%
\bibitem [{\citenamefont {Shargel}(2010)}]{shargel}%
  \BibitemOpen
  \bibfield  {author} {\bibinfo {author} {\bibfnamefont {B.~H.}\ \bibnamefont
  {Shargel}},\ }\href@noop {} {\bibfield  {journal} {\bibinfo  {journal} {J.
  Phys. A: Math. Gen.}\ }\textbf {\bibinfo {volume} {43}},\ \bibinfo {pages}
  {135002} (\bibinfo {year} {2010})}\BibitemShut {NoStop}%
\bibitem [{\citenamefont {Speck}\ \emph {et~al.}(2007)\citenamefont {Speck},
  \citenamefont {Blickle}, \citenamefont {Bechinger},\ and\ \citenamefont
  {Seifert}}]{speck2007}%
  \BibitemOpen
  \bibfield  {author} {\bibinfo {author} {\bibfnamefont {T.}~\bibnamefont
  {Speck}}, \bibinfo {author} {\bibfnamefont {V.}~\bibnamefont {Blickle}},
  \bibinfo {author} {\bibfnamefont {C.}~\bibnamefont {Bechinger}}, \ and\
  \bibinfo {author} {\bibfnamefont {U.}~\bibnamefont {Seifert}},\ }\href@noop
  {} {\bibfield  {journal} {\bibinfo  {journal} {Europhys. Lett.}\ }\textbf
  {\bibinfo {volume} {79}},\ \bibinfo {pages} {30002} (\bibinfo {year}
  {2007})}\BibitemShut {NoStop}%
\end{thebibliography}
\end{document}